\documentclass[fleqn,usenatbib]{mnras}

\usepackage{newtxtext,newtxmath}

\usepackage[T1]{fontenc}
\usepackage{ae,aecompl}
\usepackage[utf8]{inputenc}
\usepackage{bm}

\usepackage{graphicx}	
\usepackage{amsmath}	
\usepackage[acronym,shortcuts]{glossaries} 
\usepackage{booktabs}
\glsdisablehyper 
\usepackage{ulem} 
\usepackage{xspace}
\usepackage[dvipsnames]{xcolor} 
\usepackage{subfig}

\newcommand{\Rsun}{\ensuremath{\,\rm{R}_{\odot}}\xspace}

\newcommand{\Msun}{\ensuremath{\,\rm{M}_{\odot}}\xspace}

\newcommand{\Lsun}{\ensuremath{\,\rm{L}_{\odot}}\xspace}

\newcommand{\Phantom}{{\scshape phantom}\xspace }
\newcommand{\SPLASH}{{\scshape splash}\xspace }
\newcommand{\MESA}{{\scshape mesa}\xspace }
\newcommand{\HII}{H\,{\scshape ii}\xspace }
\newcommand{\HI}{H\,{\scshape i}\xspace }
\newcommand{\HeIII}{He\,{\scshape iii}\xspace }
\newcommand{\HeII}{He\,{\scshape ii}\xspace }
\newcommand{\HeI}{He\,{\scshape i}\xspace }

\newacronym{RSG}{RSG}{red supergiant}

\newacronym{CE}{CE}{common-envelope}
\newacronym{RLOF}{RLOF}{Roche-lobe overflow}
\newacronym[longplural={supernovae}, shortplural={SNe}]{SN}{SN}{supernova}

\newacronym{SPH}{SPH}{smoothed particle hydrodynamics}
\newacronym[longplural={equations of state}, plural={EoSs}]{EoS}{EoS}{equation of state}

\title[Common envelopes in massive stars]{Common envelopes in massive stars:  Towards the role of radiation pressure and recombination energy in ejecting red supergiant envelopes}

\author[M. Y. M. Lau et al.]{Mike Y. M. Lau,$^{1,2}$\thanks{E-mail: mike.lau@monash.edu (MYML)}
	Ryosuke Hirai$^{1,2}$,
	Miguel Gonz\'{a}lez-Bol\'{i}var$^{3,4}$,
	\newauthor
	Daniel J. Price$^{1}$,
	Orsola De Marco$^{3,4}$,
	and Ilya Mandel$^{1,2,5}$
	\\
	$^{1}$ School of Physics and Astronomy, Monash University, Clayton, Victoria 3800, Australia\\
	$^{2}$ OzGrav: The ARC Centre of Excellence for Gravitational Wave Discovery, Australia\\
	$^3$ Department of Physics and Astronomy, Macquarie University, Sydney, NSW 2109, Australia\\
    $^4$ Astronomy, Astrophysics and Astrophotonics Research Centre, Macquarie University, Sydney, NSW 2109, Australia\\
    $^5$ Birmingham Institute for Gravitational Wave Astronomy and School of Physics and Astronomy, University of Birmingham, \\Birmingham, B15 2TT, United Kingdom
}

\date{Accepted XXX. Received YYY; in original form ZZZ}

\pubyear{2021}

\begin{document}
\label{firstpage}
\pagerange{\pageref{firstpage}--\pageref{lastpage}}
\maketitle

\begin{abstract}
	We perform 3D hydrodynamical simulations of a common-envelope event involving a 12\Msun red supergiant donor. Massive stars are expected to be qualitatively different from low-mass stars as their envelopes have significant support from radiation pressure, which increases both the final separation and amount of mass ejected through the common-envelope interaction. We perform adiabatic simulations that include radiation energy through the equation of state, which results in ejecting 60 per cent more mass (up to two thirds of the total envelope mass becoming unbound, or more) and yield a 10 per cent larger final separation compared to simulations that assume an ideal gas. When also including recombination energy, we find that at least three quarters of the envelope, and possibly the entire envelope, may be unbound. The final separation further increases by almost 20 per cent. The additional amount of ejected material is mainly due to energy injected from helium recombination. Hydrogen recombination plays a comparatively small role, as it mainly occurs in gas that has already become unbound. We conclude that the internal energy of the envelope can be a significant energy source for ejecting the common envelope, but ultimately radiation transport and convection need to be included.
\end{abstract}



\section{Introduction}
A \ac{CE} event is a short-lived but transformational phase that occurs when a companion star in a binary enters the extended envelope of a giant star (the `donor') and orbits the giant's core inside \citep{Paczynski1976,Ivanova+2013}. Despite being a critical process in determining the fate of a binary star, \ac{CE} evolution is one of the least understood phases in binary evolution. \ac{CE} evolution has been studied extensively over the past few decades using a range of approaches, including global hydrodynamical simulations \citep{Rasio&Livio1996,Sandquist+1998,Ricker&Taam2008,Passy+2012,Ricker&Taam2012,Nandez+14,Ohlmann+16,Ivanova&Nandez16,Iaconi+2017,Chamandy+2018,Reichardt+2019,Prust&Change19,Kramer+20,Sand+20,Glanz+21}, local wind-tunnel simulations \citep{MacLeod&Ramirez-Ruiz2015,MacLeod+2017,De+20}, and 1D simulations \citep{Ivanova+15,Clayton:2017,Fragos+2019}.

To date, global 3D simulations have not demonstrated complete envelope ejection without additional energy sources such as recombination. Limited by resolution or energy non-conservation, many simulations also do not model the \ac{CE} phase from the onset of \ac{RLOF}, and instead initiate the companion on or close to the surface of an unperturbed donor star \citep[but see][]{Iaconi+2017,Reichardt+2019}. Extensive resolution studies are scarce due to long computation times, and, where performed, show that some key quantities fail to converge.

Moreover, previous global simulations have focused almost entirely on \acp{CE} involving low mass donors (1--2\Msun). In this study, we focus instead on the \ac{CE} evolution of massive stars, which is considered a standard scenario for forming the close binary progenitors of a range of astrophysical phenomena, including low- and high-mass X-ray binaries \citep[see][and references therein]{Kalogera1998,vandenHeuvel2019}, double neutron stars \citep[e.g.][]{Tauris+2017,Vigna-Gomez+2018}, stripped-envelope supernovae \citep[e.g.][]{Podsiadlowski+92,Yoon15}, and gravitational waves emitted during compact binary coalescences \citep[see][and references therein]{Mandel+Broekgaarden21}.

We expect \ac{CE} evolution to be qualitatively different for massive stars, since their envelopes have significant pressure support from radiation. So, unlike low-mass donor stars, radiation pressure must be included in their modelling. \cite{Taam+78} performed an early 1D study of a massive star \ac{CE} involving a 16\Msun donor and a 1\Msun neutron star companion, which was later extended to 2D \citep{Bodenheimer+Taam84} and 3D \citep{Terman+95}. Recently, \cite{Fragos+2019} performed a 1D simulation involving a 12\Msun donor in the context of forming merging double neutron stars. \cite{Law-Smith+20} studied a 12\Msun donor with 3D hydrodynamics, but only simulated the last 8\Rsun of the dynamical inspiral. \cite{Ricker+2018} presented preliminary results from a 3D simulation that included flux-limited radiation diffusion with a 82.1\Msun donor and 35\Msun black hole companion. While this paper was under review, the manuscript by \cite{Moreno+21} became available, which presented \ac{CE} simulations involving a 10\Msun donor with a 1.4\Msun or a 5\Msun companion.

In this paper, we present results from a set of 3D global \ac{CE} simulations involving a 12\Msun \ac{RSG} donor. We perform the same simulation (fixed initial separation, donor and companion masses, and hydrostatic profile of the donor star) with different \acp{EoS}: (i) Ideal gas, (ii) Gas + radiation, and (iii) `Full' \ac{EoS}, which, in the order listed, sequentially include the pressure and/or internal energy contributed by gas, radiation, and recombination. Comparing simulations performed with the different \acp{EoS} will reveal the role of internal energy in ejecting the \ac{CE}, particularly radiation thermal energy, which is only significant in massive stars.

Like almost all \ac{CE} hydrodynamical simulations that have been performed to date, our simulations are adiabatic, and so do not include radiation transport and envelope convection. Efficiently transporting energy through radiation or convection could reduce the ability of orbital and recombination energy to do work on the envelope. The retention of recombination energy in the envelope, particularly hydrogen recombination energy, is unclear \citep{Sabach+17,Grichener+18,Soker+18}. Simulations including radiation transport are needed to assess how much recombination energy would be transported away by radiation and/or convection. Our adiabatic simulations therefore place upper limits on the efficiency of various energy sources (see Section \ref{subsec:energy_transport}).

This paper is organised as follows. In Section \ref{sec:methods}, we summarise our simulations (\ref{subsec:overview}), then describe our donor star (\ref{subsec:donor}), initial conditions (\ref{subsec:initial}), and the \acp{EoS} used across different simulations (\ref{subsec:eos}). In Section \ref{sec:results}, we present our main results, including the final separation (\ref{subsec:sep}) and amount of unbound envelope mass (\ref{subsec:unbinding}), and verify energy and angular momentum conservation (\ref{sec:conservation}). In Section \ref{sec:discussion}, we discuss the implications of our results, including the mass-loss mechanism and role of recombination (\ref{subsec:unbinding_mech}), the conditions surrounding the termination of the dynamical plunge-in (\ref{subsec:termination}), values of the \ac{CE} efficiency parameter $\alpha$ inferred from the simulations (\ref{subsec:alpha}), the importance of modelling energy transport in massive star \ac{CE} simulations (\ref{subsec:energy_transport}), and astrophysical implications (\ref{subsec:implications}). We summarise our findings in Section \ref{sec:conclusion}.

\section{Methods} \label{sec:methods}

\subsection{Overview} \label{subsec:overview}
We use the \ac{SPH} code, \Phantom \citep[v2021.0.0,][]{Price+18}, to perform simulations involving a 12\Msun \ac{RSG} donor. We include a total of 18 different runs that differ in the companion mass ($M_2$), assumed \ac{EoS}, and/or numerical resolution (see Table \ref{tab:summary} for a summary of the highest resolution simulations). Half of the runs are carried out with a 1.26\Msun companion ($q = 0.11$), while the other half are carried out with a 3\Msun companion ($q = 0.25$). 

\Phantom has been applied to \ac{CE} simulations involving low-mass donors \citep[e.g.,][]{Iaconi+2017,Reichardt+2019,Reichardt+20}. It uses a state-of-the-art shock-capturing scheme based on a modification of the \cite{Cullen+Dehnen10} shock detector as described in Section 2.2 of \cite{Price+18}. Previous studies have compared \ac{SPH} and grid-based codes for \ac{CE} simulations. \cite{Passy+2012} compared simulations involving a 0.88\Msun red giant branch donor performed by \textsc{SNSPH} and by the grid-based code \textsc{Enzo}. \cite{Iaconi+2017} compared simulations performed with \Phantom and with \textsc{Enzo} for the same donor used in \cite{Passy+2012}. Both studies found agreement between the final separation and amount of unbound mass at a 10\% level.

We perform simulations with three different \acp{EoS} (see Section \ref{subsec:eos}): (i) Ideal gas; (ii) Gas + radiation; and (iii) `Full'. The ideal gas \ac{EoS} assumes an adiabatic index of $\gamma = 5/3$. The gas + radiation \ac{EoS} includes the pressure and thermal energy contributed by both gas and radiation; we do not perform radiation transport, and so assume local thermodynamic equilibrium. The full \ac{EoS} refers to \Phantom's implementation of the \ac{EoS} tables provided by \MESA \citep{MESA1}, which further includes recombination. Including radiation in the \ac{EoS} effectively reduces the adiabatic index, thus increasing the amount of internal energy stored in the stellar envelope compared to the case where the envelope material is treated as an ideal gas. Comparing simulations performed with the different \acp{EoS} will reveal the roles of internal energy and recombination in massive star \ac{CE} evolution. 

The resolution of a simulation is determined by the number of \ac{SPH} particles used to model the \ac{RSG}. We perform simulations with 50k (`low resolution'), 500k (`medium resolution'), and 2M (`high resolution') \ac{SPH} particles to study the convergence of our results. All results and figures presented in this paper pertain to the high-resolution simulations unless indicated otherwise. Where relevant, we compare results obtained at different resolutions, showing that our qualitative conclusions are robust against resolution. 

We perform our simulations mainly with the Gadi supercomputer's Cascade Lake nodes, each containing two 24-core Intel Xeon Platinum 8274 3.2 GHz processors, and the OzSTAR supercomputer's Sandy Bridge nodes, each containing two 8-core Intel Xeon E5-2660 processors. A single high-resolution \ac{CE} simulation requires roughly 40--50 kcpu-hr, taking roughly 2--3 months of wall time. A medium-resolution calculation, starting at a larger initial separation, has a similar resource usage. Our simulations use OpenMP parallelisation, and show nearly linear scaling between computation speed and core number up to at least 48 cores. Future optimisation for MPI parallelisation in \Phantom would allow us to use a higher resolution and to track the post-plunge-in evolution for longer. 

\begin{table}
	\centering
	\begin{tabular}{llllll}
		\toprule
		$q$ & EoS & $a_\mathrm{i}$ /\Rsun & $a_\mathrm{f}$ /\Rsun & $f_{\rm{k}+\rm{p}+\rm{th}}$ & $f_{\rm{k}+\rm{p}}$ \\ \midrule
		0.25 & Ideal gas  & 988 & 33.1    & 0.18     & 0.16     \\
		0.25 & Gas + rad. & 988 & 37.6    & 0.28     & 0.24     \\
		0.25 & Full       & 988 & 44.3    & 0.60     & 0.40     \\
		0.11 & Ideal gas  & 862 & $<22.8$ & $>0.033$ & $>0.031$ \\
		0.11 & Gas + rad. & 862 & $<22.8$ & $>0.072$ & $>0.036$ \\
		0.11 & Full   	  & 862 & $<22.8$ & $>0.10$  & $>0.054$ \\
		\bottomrule
	\end{tabular}
	\caption{Summary of main results for the high-resolution simulations. From left to right, the columns list (i) $q = M_2/M_1$, the companion to donor mass ratio; (ii) The assumed equation of state; (iii) The initial separation; (iv) The orbital separation when the ratio of the orbital period to the inspiral time-scale falls below a threshold ($-P_\text{orb}\dot{a}/a < 5\times10^{-4}$). Columns (v) and (vi) report the fraction of unbound envelope mass at that same reference point, with $f_{\rm{k}+\rm{p}}$ assuming a purely mechanical criterion for considering material to be unbound and $f_{\rm{k}+\rm{p}+\rm{th}}$ also including thermal energy (see Section \ref{subsec:unbinding}). For the $q=0.11$ simulations, the unbound mass fraction is reported at the minimum core-companion separation 22.8\Rsun, where the dynamical plunge-in is still occurring.}
	\label{tab:summary}
\end{table}
\subsection{The donor star} \label{subsec:donor}

\begin{figure}
	\centering
	\includegraphics[width=\linewidth]{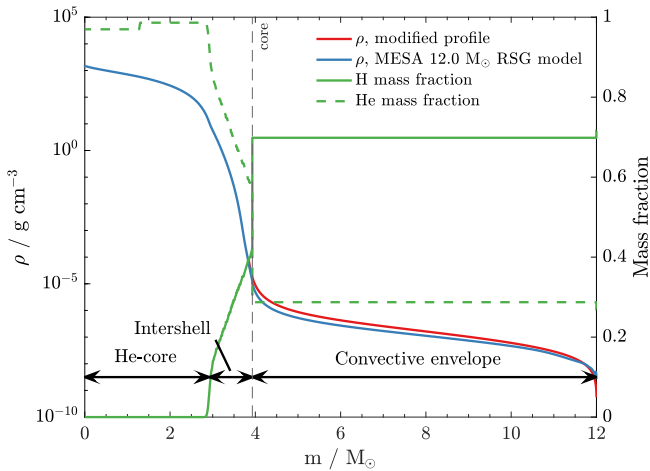}
	\caption{Comparison of the density profile of the donor star used in our simulations (red) with that of a reference 12\Msun red supergiant model (blue) evolved using \MESA to its maximum radius during core helium burning. The hydrogen and helium mass fractions (green solid and green dashed lines) are also shown, indicating the locations of the helium core, the convective envelope, and the thin intershell. Below $m =3.84\Msun$ ($r_\text{core} = 18.5\Rsun$, marked with the grey, vertical dashed line), we replace the dense core with a point mass embedded in a softened density profile.}
	\label{fig:rho_profile}
\end{figure}

\begin{figure*}
	\centering
	\includegraphics[width=\linewidth]{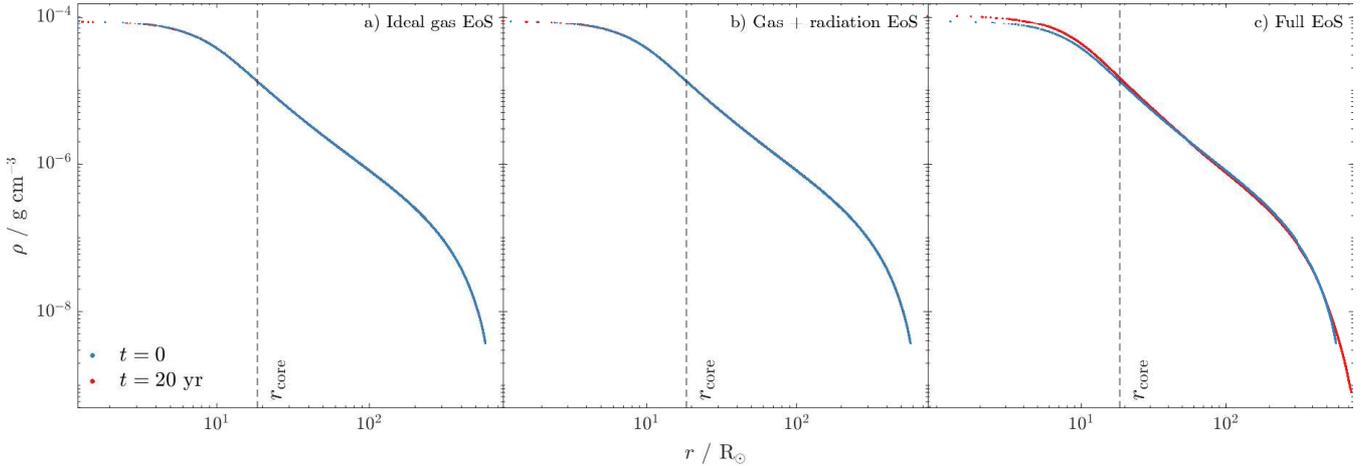}
	\caption{Comparison of the donor star density profiles immediately after our relaxation procedure ($t=0$, blue markers) and after another twenty years (red markers), which is 89 times the surface free-fall time. This demonstrates that the star evolved in isolation is stable. The left, centre, and right panels correspond to different equations of state assumed while evolving the single star: a) Ideal gas, b) Gas + radiation, and c) Full \ac{EoS}.}
	\label{fig:stability}
\end{figure*}

\acp{RSG} are known to have deep convective envelopes. Convection is driven by a negative entropy gradient sustained with central heating from nuclear burning and surface cooling from radiative losses (see Section \ref{subsec:energy_transport}). As a first attempt, we do not model these effects for simplicity, and perform fully adiabatic simulations. We create an artificial, convectively stable \ac{RSG} model by solving the hydrostatic equations while assuming uniform entropy throughout the envelope (see Appendix \ref{app:shooting} for details), and specify the stellar radius and core size as boundary conditions. 

We choose the characteristic radius and core size of the donor based on a 12\Msun stellar model evolved with 1D stellar evolution code \MESA \citep[v12115,][]{MESA1,MESA2,MESA3,MESA4,MESA5}\footnote{We use an initial metallicity of $Z=0.0142$ and the default settings in \texttt{inlist\_massive\_defaults}, but with step overshooting above the hydrogen-burning core, hydrogen-burning shell, and helium-burning core by 0.2 times the pressure scale height near the top of the convective boundary. We do not include wind mass loss.} to its maximum radius during core helium burning. We choose this model as it has the lowest envelope binding energy within the core-helium burning stage, therefore plausibly easing envelope ejection. On the other hand, choosing an earlier, more tightly bound model may decrease the post-\ac{CE} separation and amount of unbound mass. A donor earlier in its evolution would also have a hotter and more radiation-dominated envelope, and so is likely to yield greater differences between the ideal gas \ac{EoS} and gas + radiation \ac{EoS} simulations. The donor mass was selected towards the lower end of the massive star range, and therefore represents a relatively abundant donor choice. This reference model has a 619\Rsun radius and a luminosity of $4.4\times10^4$ L${}_\odot$.

The location of the bifurcation point between the ejected envelope and the remaining core in a \ac{CE} is uncertain. We choose to replace the region below the convective part of the envelope, located at $r_\text{core} = 18.5\Rsun$ ($m_\text{core} = 3.84\Msun$), with a point mass due to computational limitations (see Appendix \ref{app:softening} for details). If we evacuate the convective part of the envelope, the remaining radiative layers are expected to expand while reestablishing thermal equilibrium. Depending on the post-\ac{CE} separation, the donor may overflow its Roche lobe and enter thermal time-scale ($\sim10^3$ yr) mass transfer \citep[see][ and Section \ref{subsec:termination}]{Vigna-Gomez+21}. Our simulations inform us of the separation at the end of the rapid inspiral as long as the donor core does not overflow its Roche lobe at the end of the rapid inspiral. Our assumed core is significantly larger than what is typically assumed to be the post-\ac{CE} remnant in binary evolution studies. As a comparison, the helium core size\footnote{We define the helium core as the region having hydrogen mass fraction below $10^{-6}$.} of the \MESA model is 0.25\Rsun (2.8\Msun), where the stellar material is much more tightly bound. The 1.0\Msun of material between the helium core and our chosen core-envelope boundary contains $2.1 \times 10^{49}$ erg of gravitational binding energy; this is 20 times the binding energy of the entire region we simulate. However, \cite{Moreno+21} explored the effects of different core-envelope boundaries in their \ac{CE} simulations involving a 10\Msun \ac{RSG} donor. They find that the final separations and amount of ejected material are not significantly affected by their choice, even when placing the boundary slightly above the base of the convective part of the envelope. We caution, however, that \cite{Moreno+21} used a different code and a different donor star model, and so their finding may not directly apply to our simulations.

Fig. \ref{fig:rho_profile} compares the density profile of the \ac{RSG} donor used in our simulations (red line) with that of the \MESA \ac{RSG} model (blue line). The vertical dashed line marks the core-envelope boundary. Despite using an artificial entropy profile, the density structure of the envelope closely follows that of the \MESA \ac{RSG} model. The artificial envelope is slightly more strongly bound than the \MESA model's envelope, with at most 10 per cent difference in gravitational binding energy. Therefore, our progenitor choice is expected to underestimate the post-\ac{CE} orbital separation.

\subsubsection{Mapping the 1D stellar profile to a 3D star}
We map and relax the 1D donor profile into a 3D distribution of \ac{SPH} particles using a new procedure we have implemented in \Phantom, which we describe in detail in Appendix \ref{app:asynchronous}. The core idea of the procedure is to shift the positions of \ac{SPH} particles iteratively while fixing the density-pressure relation, which is guaranteed to accurately reproduce the intended density profile. To make this relaxation process more efficient, we shift the particles \textit{asynchronously}, according to their local stable time-step.

Fig. \ref{fig:stability} and \ref{fig:donor_profiles} demonstrate the stability and accuracy of the star produced from our mapping procedure. The top panel of Fig. \ref{fig:donor_profiles} shows that the relaxed donor star (green markers) is nearly identical to the original stellar profile (black dashed line). In the absence of a companion, the star maintains its original density profile to a high degree. Fig. \ref{fig:stability} compares the density profile of the \ac{RSG} donor immediately after relaxation (blue) and after another 20 years (red), which is approximately 89 times the surface free-fall time. With the ideal gas \ac{EoS} (Fig. \ref{fig:stability}(a)) and gas + radiation \ac{EoS} (Fig. \ref{fig:stability}(b)), there are barely any noticeable deviations. To quantify this, we calculate the fractional difference between the mass interior to radius $r$ in the 3D star that was evolved for 20 years and the mass coordinate $m(r)$ of the 1D profile. For the donor evolved with the ideal gas \ac{EoS}, the maximum difference is about a few times $10^{-4}$, whereas with the gas + radiation \ac{EoS}, it is about $10^{-3}$. There are more noticeable deviations with the full \ac{EoS} (Fig. \ref{fig:stability}(c)), where the star relaxes into an equilibrium configuration with a 25 per cent larger radius and 15 per cent greater central density (see Section \ref{subsec:energy_transport}). These results are obtained at the lowest resolution (50k particles); we find improved stability at higher resolutions.

\subsection{Initial conditions} \label{subsec:initial}
We set up the binary to be in a circular orbit, and assume the donor star to be non-spinning and unperturbed by its companion in its previous evolution up to this point. While this is not the most realistic setup, there are many uncertain processes leading to the start of the \ac{CE} phase that may modify the angular momentum and envelope structure of a donor star, including wind mass loss, tidal interactions, and thermal time-scale mass transfer. The aim of our current study is not to model all of these effects, but to quantify the impact of the different \acp{EoS}. As such, we choose initial conditions that enable a clean comparison between models rather than aiming to create the most astrophysically accurate setup.

For the low- and medium-resolution simulations, we chose the initial separation, $a_i$, such that the donor star fills its Roche lobe, whose volumetric radius is evaluated using the expression by \cite{Eggleton1983}. For the $q=0.25$ ($M_2 = 3\Msun$) case, this is $a_i = 1234\Rsun$. Whereas for the $q=0.11$ ($M_2 = 1.26\Msun$) case, this is $a_i = 1078\Rsun$. Typically, the companion makes many orbits before plunging into the donor envelope. At high resolution, the wall time required to simulate this phase is unfeasibly long. We therefore start the simulations with a smaller initial separation, such that the stellar radius exceeds the Roche lobe by 25 per cent. This translates to $a_i = 988\Rsun$ for the $q=0.25$ simulations, and $a_i = 862\Rsun$ for the $q=0.11$ simulations. In Appendix \ref{app:roche-filling}, we demonstrate that this smaller initial separation does not significantly alter the post-\ac{CE} orbital separation.
\subsection{Equation of state} 
\label{subsec:eos}

\begin{figure}
	\centering
	\includegraphics[width=\linewidth]{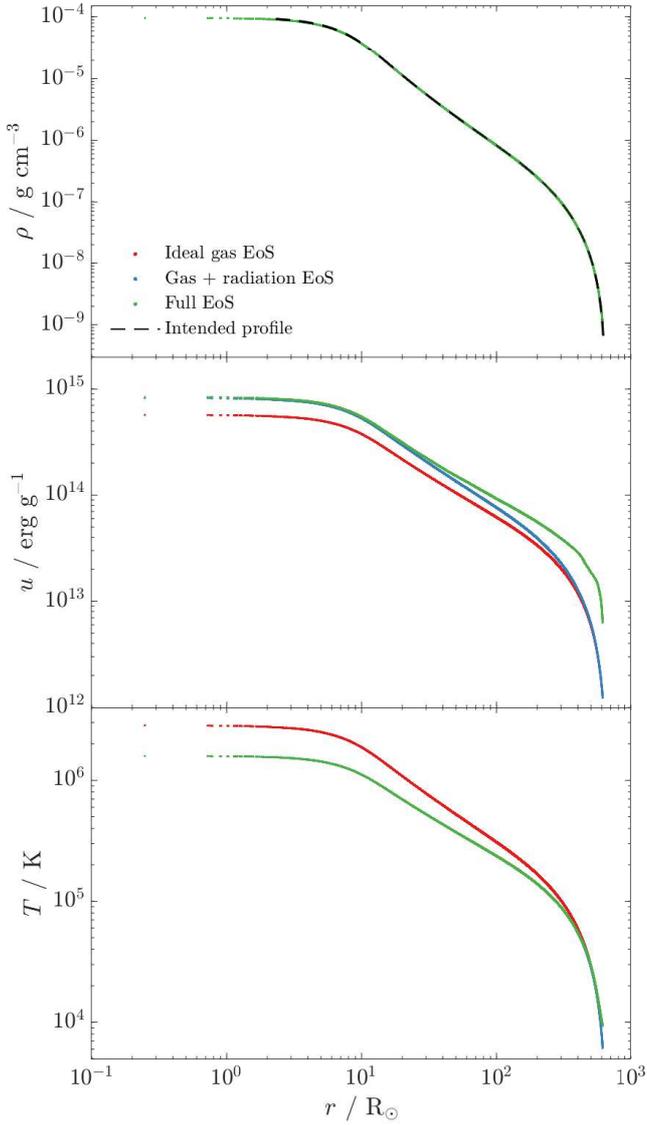}
	\caption{Profiles of the red supergiant donor at $t=0$ in our simulations (directly after relaxation) for different assumed equations of state: (i) Ideal gas (red), (ii) Gas + radiation (blue), and (iii) Full equation of state (green). Each marker corresponds to a single \ac{SPH} particle in a red supergiant resolved with two million \ac{SPH} particles. \textit{Top panel}: The density profile, $\rho(r)$, which is independent of the assumed \ac{EoS}. The relaxed star is in excellent agreement with the intended 1D profile (black dashed curve). \textit{Middle panel}: The specific internal energy profile, $u(r)$. \textit{Bottom panel}: The temperature profile, $T(r)$.} 
	\label{fig:donor_profiles}
\end{figure}

\begin{figure}
	\includegraphics[width=\linewidth]{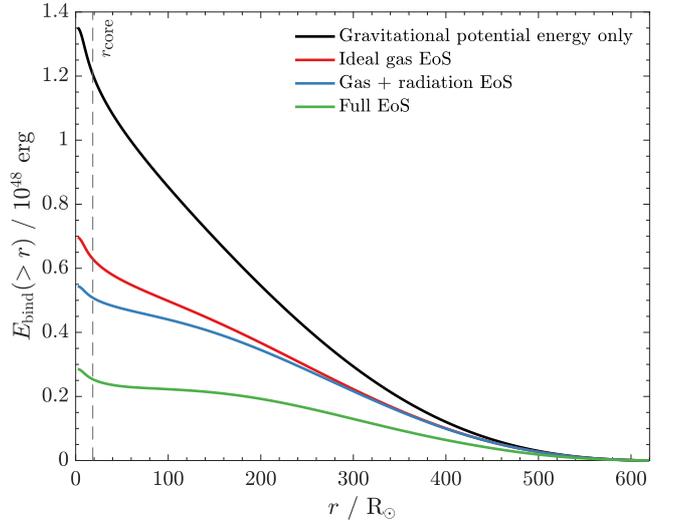}
	\caption{Comparison of the cumulative binding energy, $E_\text{bind}(>r)$ (defined in equation (\ref{eq:ebind})), integrated from the surface. The black curve is the gravitational binding energy while the coloured curves include various forms of internal energy depending on the assumed equation of state: Ideal gas (red), gas + radiation (blue), and full equation of state (green). The gravitational potential in $r<r_\text{core}$ is softened (see Appendix \ref{app:softening}).}
	\label{fig:ebind_profile}
\end{figure}

We carry out the same \ac{CE} simulation with three different \acp{EoS}: (i) Ideal gas, (ii) Gas + radiation, and (iii) `Full'. In the order listed, these \acp{EoS} progressively include the effects of gas pressure, radiation pressure, and recombination. Existing \ac{CE} simulations, which mainly target low-mass stars, typically assume an ideal gas \ac{EoS}, which is adequate for modelling the gas-pressure-dominated envelopes of low-mass stars. For the massive stars that we wish to simulate, radiation pressure is non-negligible, and could comprise a substantial fraction of the envelope binding energy. Comparing simulations assuming an ideal gas \ac{EoS} with those assuming a gas + radiation \ac{EoS} therefore reveals the qualitative difference between a low-mass and a high-mass \ac{CE}.

In addition, recombination energy in giant envelopes could serve as a significant extra energy source for envelope ejection \citep[][but see \citealt{Sabach+17,Soker+18,Grichener+18}]{Ivanova+15,Ivanova&Nandez16,Ivanova18,Reichardt+20}. However, the efficiency of various forms of recombination is not fully understood, and there has been no global hydrodynamical simulation exploring the role of recombination in massive stars. Comparing our simulations that assume a gas + radiation \ac{EoS} with those that assume the full \ac{EoS} therefore reveals the role of recombination energy in ejecting the envelope.

For the ideal gas \ac{EoS}, we use an adiabatic exponent $\gamma = 5/3$, corresponding to a classical monatomic gas. For the gas + radiation \ac{EoS}, we also include radiation pressure and internal energy, assuming local thermodynamic equilibrium and without radiation transport:
\begin{align}
	P(\rho,T) &= \frac{\rho k_\mathrm{B}T}{\mu m_\mathrm{H}} + \frac{1}{3}a_\mathrm{rad}T^4,
	\label{eq:pres} \\
	u(\rho,T) &= \frac{3 k_\mathrm{B}T}{2\mu m_\mathrm{H}} + \frac{a_\mathrm{rad}T^4}{\rho},
	\label{eq:ene}
\end{align}
where $P$, $\rho$, $T$, $u$ are the pressure, density, temperature, and specific internal energy. $\mu$ is the mean molecular weight, $k_\mathrm{B}$ is the Boltzmann constant, $m_\mathrm{H}$ is the hydrogen mass, and $a_\mathrm{rad}$ is the radiation constant. We set a uniform $\mu = 0.62$, which corresponds to a hydrogen mass fraction of $X = 0.70$ and metallicity of $Z=0.014$. For the full \ac{EoS}, we use \Phantom's implementation of the \ac{EoS} tables provided by \MESA as described in \cite{Reichardt+20}, which combine the OPAL \ac{EoS} \citep{Rogers+96,Rogers+Nayfonov02} and SCVH \ac{EoS} \citep{Saumon+95}. The full \ac{EoS} includes recombination energy, which is dominated by helium and hydrogen (including molecular recombination into H${}_2$). The mean molecular weight is a function of the ionisation fractions of all species, and is outputted from the \ac{EoS} table rather than prescribed to the \ac{EoS}.

\subsubsection{Impact of equation of state on the initial donor profile}
\label{subsubsec:donor_eos}
We construct the density and pressure profiles of the initial donor star to be the same across all \acp{EoS}. Using different \acp{EoS} therefore only alters the thermal profile (temperature and internal energy). This is shown in Fig. \ref{fig:donor_profiles}, which shows the density, internal energy, and temperature profiles of the \ac{RSG} donor at $t=0$ in our simulations. Stellar profiles obtained from different assumed \acp{EoS} are plotted with different marker colours, with each marker corresponding to a single \ac{SPH} particle. In the top panel, all the profiles perfectly overlap as the density is the same for all \ac{EoS} choices. In the middle panel, we see that the internal energy computed with the gas + radiation \ac{EoS} (in blue) is larger than that computed with the ideal gas \ac{EoS} (in red), as including radiation lowers the effective adiabatic index. This difference approaches zero towards the surface, which is completely gas-dominated. The internal energy is further increased when calculated with the full \ac{EoS}, due to the inclusion of ionisation energy. The bottom panel shows that the ideal gas \ac{EoS} implies a higher temperature relative to the gas + radiation \ac{EoS} except at the gas-dominated surface. The temperature calculated with the full \ac{EoS} is equal to that calculated with the gas + radiation \ac{EoS} except near the partially-ionised surface layers.

Some fraction of the donor star's initial internal energy content may be used to unbind the \ac{CE}. This can be incorporated into an effective binding energy expression that includes internal energy. The binding energy above a radius $r$ can be defined as 
\begin{align}
	E_\text{bind}(>r) = \int_{M}^{m(r)} \bigg(
	u(r) - \frac{Gm'}{r}
	\bigg)dm',
	\label{eq:ebind}
\end{align}
where $m$ is the mass coordinate, ranging from 0 to the total mass $M$. The second term in the integrand expresses the gravitational potential energy, which only depends on the mass profile of the donor star and is therefore independent of \ac{EoS}. The first term gives the contribution from internal energy, which acts to decrease $E_\text{bind}(>r)$. We plot this expression in Fig. \ref{fig:ebind_profile} for different \acp{EoS}. The intersection of each curve with the core radius (vertical dashed line) as defined in Section \ref{subsec:donor} gives the value of the envelope binding energy. The black curve gives the gravitational binding energy integrated from the surface. The total gravitational binding energy of the envelope is $1.2 \times 10^{48}$ erg, independent of \ac{EoS}. The coloured lines include the contribution by internal energy, with the red, blue, and green curves corresponding to the ideal gas, gas + radiation, and full \ac{EoS}, respectively. The corresponding envelope binding energies are $6.3\times 10^{47}$, $5.1\times 10^{47}$, and $2.5\times 10^{47}$ erg. This means that if internal energy is fully efficient in ejecting the envelope and if the \ac{CE} is completely unbound, including radiation energy allows the orbit to stall at a 24 per cent larger final separation, while adding recombination on top of radiation allows the final orbit to stall at a further two times larger separation. Our simulations may therefore determine the efficiency of internal energy in ejecting the envelope, which is generally unclear as the internal energy and gravitational energy are tightly related via the virial theorem.

\section{Results} \label{sec:results}

\begin{figure*}
	\centering
	\includegraphics[width=\linewidth]{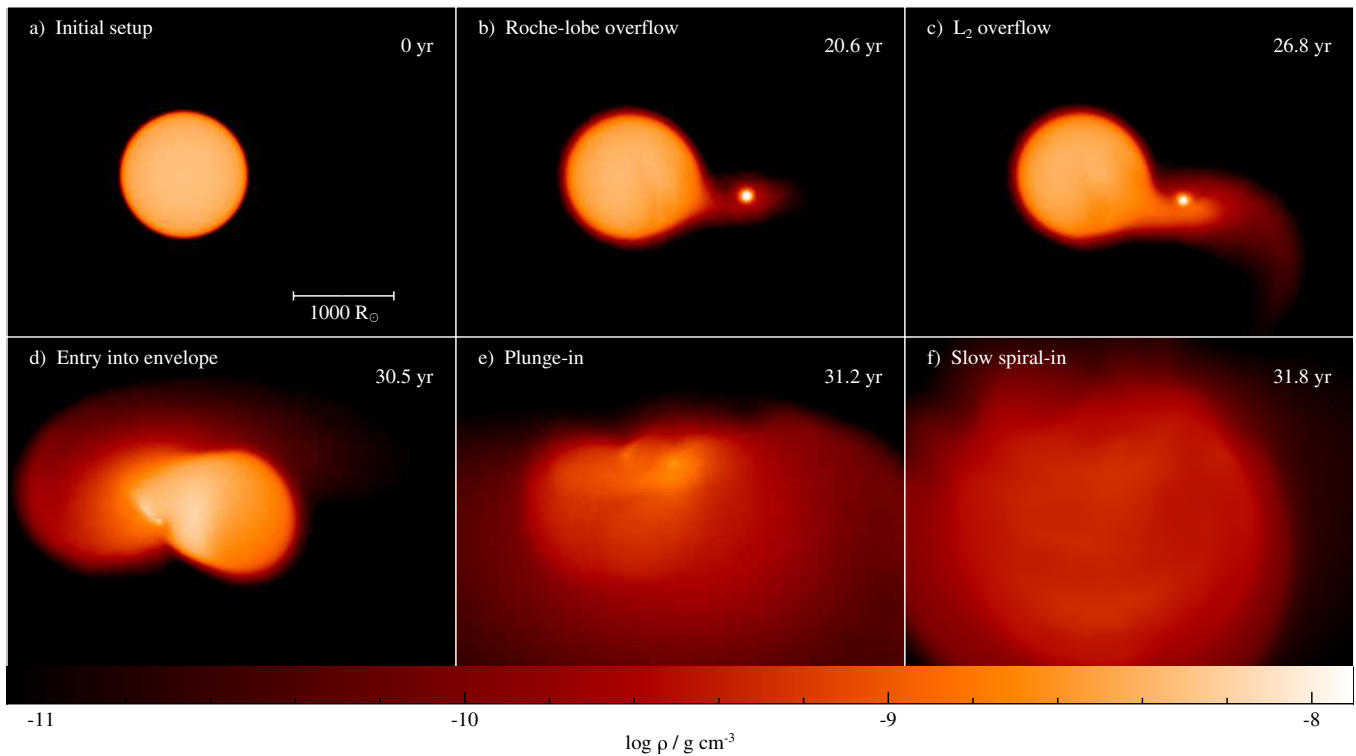}
	\caption{Surface rendering of a common-envelope simulation using ray tracing of the density field. The density at an artificial photosphere corresponding to a constant, artificially low opacity of $\kappa = 8.4\times 10^{-5}$ $\rm{cm}^{2}\rm{ g}^{-1}$ is shown for visualisation purposes. This Figure displays the medium-resolution, $q=0.25$ simulation assuming the gas + radiation equation of state, with phases a) - f) described in Section \ref{sec:results}. Videos of our simulations are available at \url{https://themikelau.github.io/RSG_CE}. All simulation renderings in this paper are produced with \SPLASH \citep{Price07}.}
	\label{fig:rho_surface_tiled}
\end{figure*}

\begin{figure*}
	\centering
	\includegraphics[width=\linewidth]{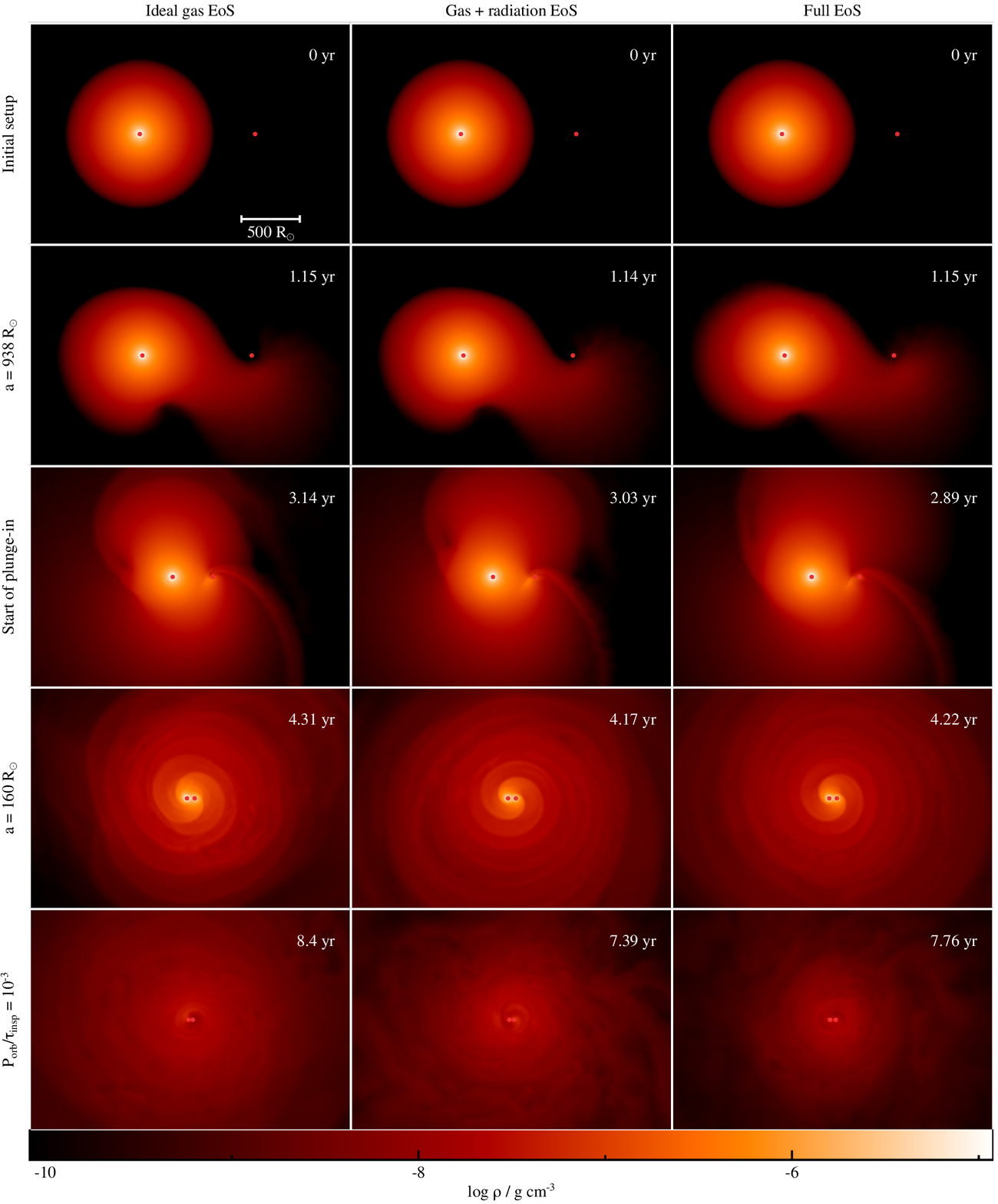}
	\caption{Each 2750-by-1896\Rsun panel shows an equatorial ($z=0$) density cross-section in a frame that is co-rotating with the orbit. The locations of the donor core and companion are marked by red dots. Different columns correspond to runs made with different equations of state, while different rows correspond to different simulation stages as labelled on the left. Row 1: Initial setup at $t=0$; Row 2: Core-companion separation reaches 938\Rsun; Row 3: Companion enters the donor envelope, marking the start of the dynamical plunge-in; Row 4: Core-companion separation is 160\Rsun; Row 5: Ratio of the orbital period of the stellar cores, $P_\text{orb}$, to the inspiral time-scale, $\tau_\text{insp} = -a/\dot{a}$, falls below $10^{-3}$. Videos of our simulations are available at \url{https://themikelau.github.io/RSG_CE}.}
	\label{fig:rho_crosssec_tiled}
\end{figure*}

\begin{figure*}
	\centering
	\includegraphics[width=\linewidth]{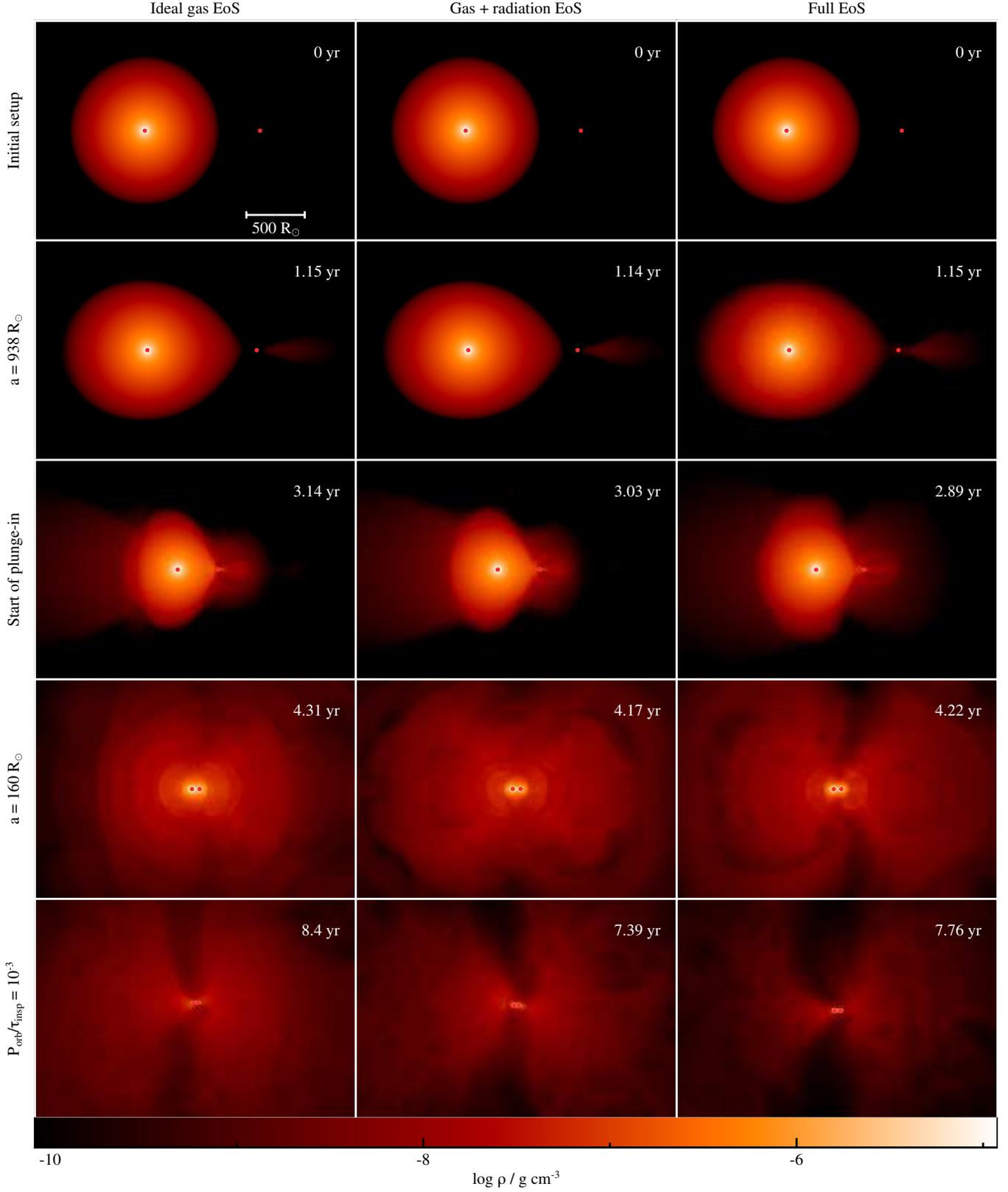}
	\caption{Same as Fig. \ref{fig:rho_crosssec_tiled}, except each 2750-by-1896\Rsun panel shows a meridional density cross-section that contains the stellar cores. Videos of our simulations are available at \url{https://themikelau.github.io/RSG_CE}.}
	\label{fig:rho_xz_tiled}
\end{figure*}

We provide an overview of the simulations in three figures. Fig. \ref{fig:rho_surface_tiled} illustrates the main phases in our \ac{CE} simulations with the $q=0.25$ medium-resolution simulation assuming the gas + radiation \ac{EoS}. Fig. \ref{fig:rho_crosssec_tiled} and \ref{fig:rho_xz_tiled} compare the simulations with different \acp{EoS} at selected moments, showing the density in the equatorial ($z=0$) and meridional cross-sections respectively at high resolution. We observe qualitatively similar evolution across different \acp{EoS} when compared at the same evolutionary phase.

Fig. \ref{fig:rho_surface_tiled}(a)-(c) show the loss of orbital stability due to \ac{RLOF}, which spans most of the simulation time, up to $\sim 100$ times the \ac{RSG} donor's surface free-fall time. For our high-resolution simulations, which start with the donor overfilling its Roche lobe by 25 per cent, this phase is much shorter, lasting only $\sim 10$ free-fall times. During this phase, the initially spherically-symmetric donor relaxes in the Roche potential, raising a tidal bulge that trails the companion's orbit. Surface gas that extends beyond the donor Roche lobe forms a $\approx 50$\Rsun radius bound disk around the companion. The disk is pressure supported and has super-Keplerian angular velocity and angular momentum. But we note that disk cooling needs to be included in order to properly simulate disk formation. In the case of a neutron star or black hole companion, there is likely to be accretion feedback, which might also occur inside the \ac{CE} \citep[e.g.,][]{Chamandy+18,Gilkis+19,Hillel+21}. Eventually, the donor overflows its $L_2$ outer Lagrangian point, rapidly driving orbital shrinkage (\ref{fig:rho_surface_tiled}(c)).

Fig. \ref{fig:rho_surface_tiled}(d) shows the companion first entering the distorted \ac{RSG} envelope, which is often considered to be the start of the \ac{CE} phase in hydrodynamical simulations. The \textit{plunge-in} phase or \textit{dynamical inspiral} ensues (Fig. \ref{fig:rho_surface_tiled}(e)), characterised by rapid orbital shrinkage. During this phase, gravitational drag dissipates orbital energy and angular momentum into the envelope, causing the companion to plunge by hundreds of solar radii over the course of several orbits. The orbital motion of the companion inside the envelope excites spiral shocks, shown in the fourth row of Fig. \ref{fig:rho_crosssec_tiled}. As this shock passes through the envelope, a modest amount of envelope gas ($\approx5$--10 per cent) becomes unbound.

Fig. \ref{fig:rho_surface_tiled}(f) shows the expanded \ac{CE} after the rapid plunge-in, where the orbit transitions to a \textit{slow spiral-in}  \citep{Meyer+Meyer-Hofmeister79,Ivanova+2013}. This phase occurs as the stellar cores spin up their surrounding gas and partially evacuate it from the orbit, significantly reducing the gravitational drag they experience (see Section \ref{subsec:termination}). We continue to simulate the evolution in this phase until the inspiral timescale is sufficiently long compared to the orbital period (as defined in Section \ref{subsec:sep}). The bottom row of Fig. \ref{fig:rho_crosssec_tiled} shows that mixing plumes develop in the \ac{CE}, while the bottom row of Fig. \ref{fig:rho_xz_tiled} shows gas being evacuated out of the orbital plane. It is unclear whether these plumes would be sustained if we allowed initial convection in the envelope.

\subsection{Final separation} \label{subsec:sep}

\begin{figure}
	\centering
	\includegraphics[width=\linewidth]{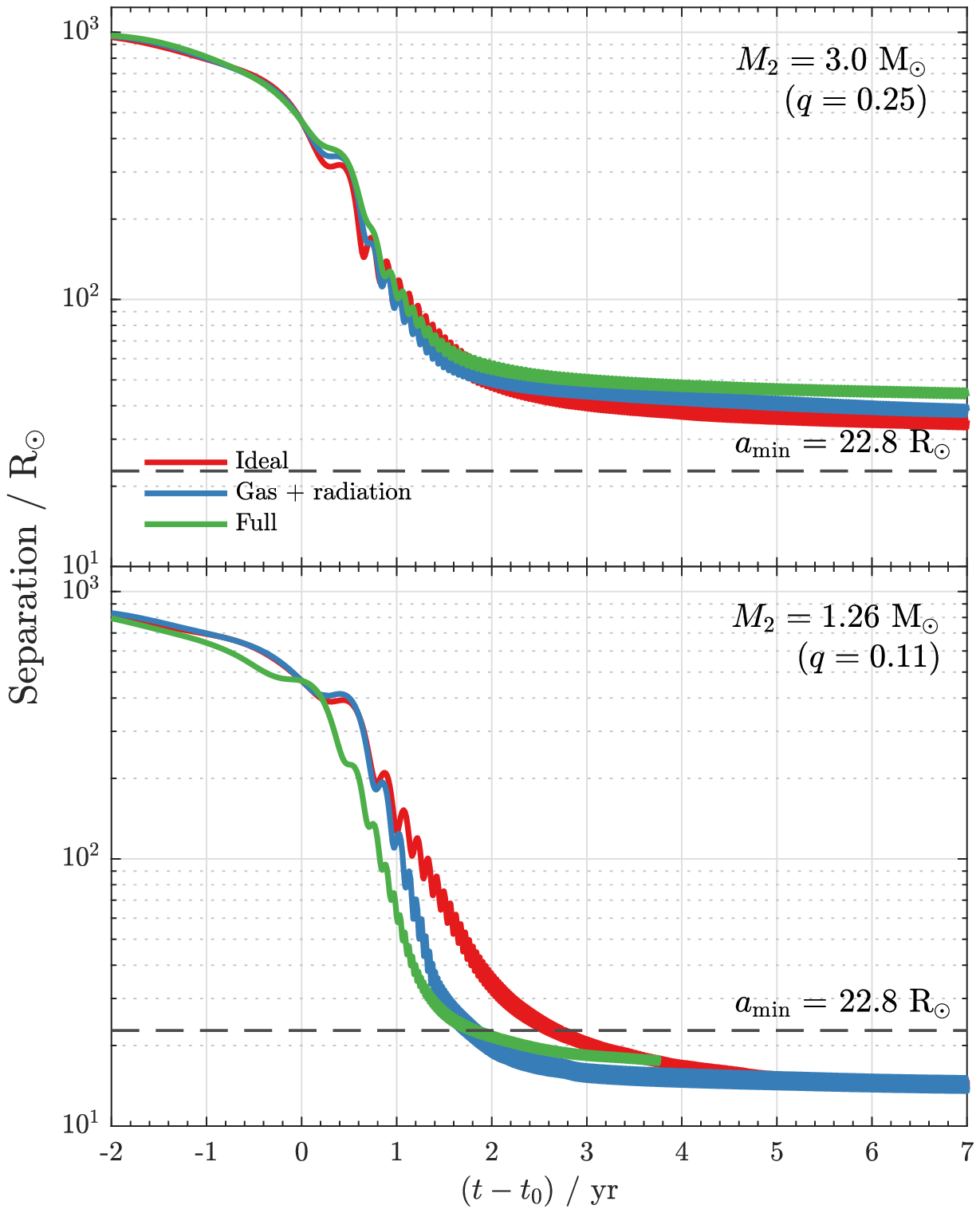}
	\caption{Separation between the companion and the donor core as a function of the time, $t-t_0$, since the core-companion separation reaches 75 per cent of the initial donor radius. Results are shown for the 3.0\Msun companion ($q=0.25$, top panel) and for the 1.26\Msun companion ($q=0.11$, bottom panel) at high resolution, and for different equations of state: Ideal gas (red), gas + radiation (blue), and the full equation of state (green). The minimum separation is 22.8\Rsun (black, horizontal dashed line), below which the final outcome of the common-envelope phase is undetermined (see Section \ref{subsec:sep}).}
	\label{fig:separation_vs_time}
\end{figure}

A central goal of \ac{CE} studies is to determine the post-\ac{CE} orbital separation and amount of envelope ejection. We study the post-\ac{CE} orbital separation of our $q=0.25$ ($M_2 = 3.0$\Msun) simulations by plotting the core-companion separation as a function of the time, $t-t_0$, since the companion separation reaches 75 per cent of the initial donor radius (top panel of Fig. \ref{fig:separation_vs_time}). We find that the post-plunge-in separation (at roughly $t-t_0\gtrsim 3$ yr) increases as we go from the ideal gas \ac{EoS} to the gas + radiation \ac{EoS} and full \ac{EoS}.

To perform a more controlled comparison, we calculate for each simulation the semi-major axis when the ratio of the inspiral time-scale to the orbital period evolves below the threshold value $-P_\text{orb}\dot{a}/a<5\times10^{-4}$. For the ideal gas, gas + radiation, and full \ac{EoS} simulations, we obtain 33.1, 37.6, and 44.3\Rsun, respectively.

These results demonstrate that the post-\ac{CE} orbital separation increases with the initial internal energy content of the envelope, as determined by the assumed \ac{EoS} (continued from Fig. \ref{fig:ebind_profile} and its discussion). Firstly, comparing the gas + radiation \ac{EoS} and the ideal gas \ac{EoS} simulations reveals that the presence of radiation pressure in massive stars causes the post-\ac{CE} orbit to stall at a larger separation, in our case, by 14 per cent, than if radiation were neglected. Including radiation increases the thermal component of the envelope internal energy by effectively decreasing the adiabatic index. Secondly, recombination energy, if thermalised locally as assumed in our adiabatic simulations, can also increase the final separation. In our case, including recombination increases the final separation by 18 per cent relative to the simulation that just includes gas and radiation energy. This finding differs from the simulation performed by \cite{Reichardt+20} using \Phantom for 0.88 and 1.88\Msun red giant donors. \cite{Reichardt+20} find that the final separation is unchanged whether assuming an ideal gas \ac{EoS} or the full \ac{EoS} (called \textit{tabulated \ac{EoS}} in their study), despite the full \ac{EoS} simulations ejecting a greater amount of envelope mass.

Compared with the higher mass ratio case, the $q = 0.11$ ($M_2=1.26$\Msun) simulations result in more orbital tightening, as shown in the lower panel of Fig. \ref{fig:separation_vs_time}. This is expected from the energy formalism, where a less massive companion has less orbital energy for a fixed semi-major axis, and so the orbit has to tighten more to extract the same amount of energy. However, we have not determined the final separations for the $q = 0.11$ simulations, as the companion reaches the minimum separation (22.8\Rsun), below which the softened region surrounding one stellar core overlaps with the other (see Appendix \ref{app:softening}). This minimum distance is plotted as a dashed horizontal line in Fig. \ref{fig:separation_vs_time}. Consequently, the outcome of the \ac{CE} phase is undetermined: the companion may stall at a deeper location inside the \ac{RSG} (i.e. beneath the convective envelope) or merge with the \ac{RSG} core (see Section \ref{subsec:termination}). Future work will aim to simulate the full plunge-in of these lower mass ratio binaries by resolving deeper layers of the donor star.

To study how well the final separation and overall rate of the plunge-in have converged across different resolutions, we plot the core-companion separation at different resolutions in the lower panels of Fig. \ref{fig:unbound_mass} for the $q=0.25$ simulations (though the high-resolution simulations have a 20 per cent smaller initial separation). We observe only minor differences in the overall rate of the plunge-in. At higher resolutions, the companion spirals in slightly faster, but also stalls at a slightly larger separation. The final separations obtained from simulations with different resolutions are listed in Table \ref{tab:resolution}, suggesting that the effect of finite resolution is to slightly underestimate the final separation. This is opposite to the trend reported by \cite{Iaconi+2017}, who also conducted a resolution test with \Phantom but using a 0.88\Msun red giant branch donor.

\begin{table}
	\centering
	\begin{tabular}{@{}lllll@{}}
		\toprule
		$N_\text{particles}$ & Equation of state & $a_\text{f}$/\Rsun & $f_{\rm{k}+\rm{p}+\rm{th}}$ & $f_{\rm{k}+\rm{p}}$ \\ \midrule
		2M    & Ideal gas 	    & 33.1 & 0.18 & 0.16 \\
		500k  & Ideal gas 	    & 34.3 & 0.19 & 0.16 \\
		50k   & Ideal gas 	    & 30.2 & 0.29 & 0.27 \\
		2M    & Gas + radiation & 37.6 & 0.28 & 0.24 \\
		500k  & Gas + radiation & 38.5 & 0.32 & 0.29 \\
		50k   & Gas + radiation & 36.0 & 0.37 & 0.31 \\
		2M    & Full            & 44.3 & 0.60 & 0.40 \\
	    500k  & Full 	        & 40.9 & 0.76 & 0.55 \\
		50k   & Full     	    & 41.0 & 0.76 & 0.54 \\
		\bottomrule
	\end{tabular}
	\caption{Comparison of key simulation quantities across different resolutions for the $q = 0.25$ ($M_2 = 3.0\Msun$) case. The first column contains the number of \ac{SPH} particles, $N_\text{particles}$, used in each simulation. Other columns are explained in the Table \ref{tab:summary} caption.}
	\label{tab:resolution}
\end{table}
\subsection{Envelope unbinding} \label{subsec:unbinding}

\begin{figure*}
	\centering
	\includegraphics[width=\linewidth]{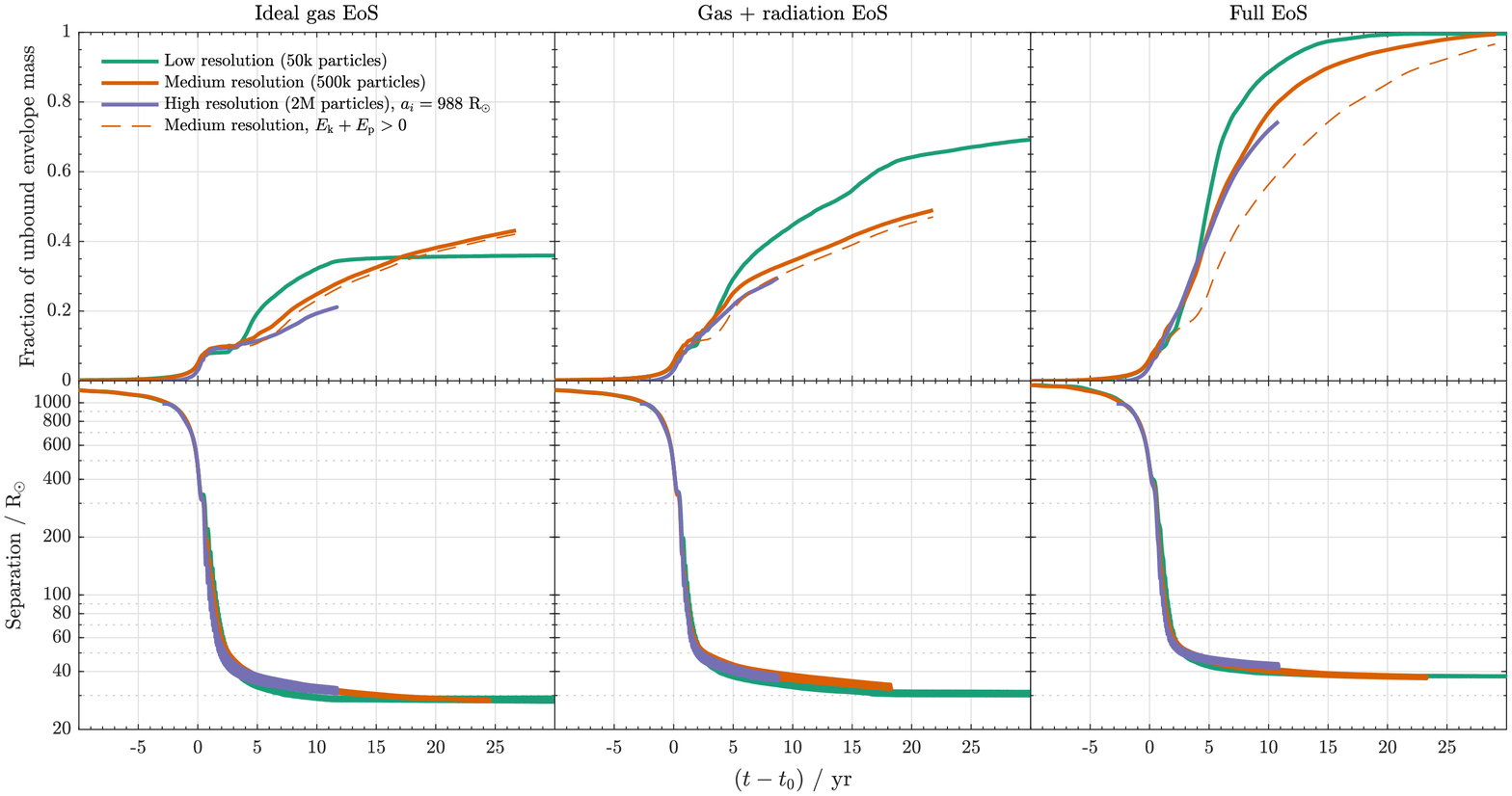}
	\caption{\textit{Top panels}: Fraction of unbound envelope mass as a function of time, $t-t_0$, since the companion separation has reached 75 per cent of the initial donor radius. \textit{Bottom panels}: Core-companion separation as a function of $t-t_0$. We show results for the $q=0.25$ simulations, with the left, centre, and right panels corresponding to the ideal gas, gas + radiation, and full \ac{EoS} cases, respectively. We plot results for different resolutions in different colours, with the low-, medium-, and high-resolution simulations shown as the green, orange, and purple lines, respectively.}
	\label{fig:unbound_mass}
\end{figure*}

Our $q=0.25$ high-resolution simulations find that at least 21-74 percent of the envelope mass is unbound depending on the assumed \ac{EoS}, with likely an even greater amount if the slow spiral-in were simulated for a longer period. Throughout this paper, we consider a single \ac{SPH} particle as unbound if the sum of its kinetic, potential, and thermal energy is positive (the thermal energy does not include recombination energy).

The top panels of Fig. \ref{fig:unbound_mass} show the fraction of unbound envelope mass in each of our $q=0.25$ simulations as a function of the shifted time, $t-t_0$. From left to right, the panels show results for the ideal gas, gas + radiation, and full \ac{EoS} simulations. The different coloured lines correspond to different resolutions as indicated by the plot legend. We can only place lower limits on the full amount of unbound envelope mass, since the unbound mass fractions for the high-resolution simulations are still increasing significantly upon terminating the simulations due to limited computational resources. For example, the amount of unbound material grows at a rate of $\approx 0.2\Msun~\mathrm{yr}^{-1}$ at the end of the full \ac{EoS} simulation. The low- and medium-resolution simulations, which have been carried out for longer, suggest that material is ejected over many years after the plunge-in.

The fraction of unbound mass has not fully converged across different resolutions. The amount of mass ejected during the plunge-in phase slightly increases with resolution\footnote{For the high-resolution case, the unbound mass curve must be shifted upwards for a fair comparison in order to correct for the smaller amount of mass unbound during the \ac{RLOF} phase, which we expect to be more similar to the medium- and low-resolution cases if the orbit were initiated at the same orbital separation.}. The differences are more significant after the plunge-in, with generally more unbound mass at lower resolutions. 

The full \ac{EoS} results in the most unbinding ($74-100$ per cent) at each resolution, followed by the gas + radiation \ac{EoS} ($>30-69$ per cent) and ideal gas \ac{EoS} ($>21-43$ per cent). As a more controlled comparison, we calculate the fraction of unbound mass, $f_{\rm{k}+\rm{p}+\rm{th}}$, at the same reference point following Section \ref{subsec:sep} ($-P_\text{orb}\dot{a}/a<5\times10^{-4}$). We find $f_{\rm{k}+\rm{p}+\rm{th}} =$ 0.18 (1.4\Msun), 0.28 (2.3\Msun), and 0.60 (4.9\Msun) for the ideal gas, gas + radiation, and full \acp{EoS}, respectively (shown in Table \ref{tab:summary}). For the $q=0.11$ simulations, Table \ref{tab:summary} instead reports $f_{\rm{k}+\rm{p}+\rm{th}}$ at the time when the companion plunges below the minimum separation (22.8\Rsun). We find that $f_{\rm{k}+\rm{p}+\rm{th}}$ follows the same trend with the assumed \ac{EoS} even though the plunge-in is still occurring.

As a comparison, we also calculate the unbound mass fraction, $f_{\rm{k}+\rm{p}}$, where we consider gas as unbound if the sum of just its specific kinetic and potential energy is positive. This criterion yields $f_{\rm{k}+\rm{p}}=$ 0.16 (1.3\Msun), 0.24 (1.9\Msun), and 0.40 (3.2\Msun) in the same order. This is a more conservative criterion as it does not account for the eventual conversion of some thermal energy into kinetic energy. We plot the evolution of $f_{\rm{k}+\rm{p}}$ for the medium-resolution simulations in Fig. \ref{fig:unbound_mass} (orange dashed line). For the ideal gas and gas + radiation \ac{EoS} simulations, $f_{\rm{k}+\rm{p}}$ and $f_{\rm{k}+\rm{p}+\rm{th}}$ are very similar. Whereas for the full \ac{EoS} simulation, the purely mechanical criterion initially marks less mass as unbound during the slow spiral-in, but eventually approaches the value of $f_{\rm{k}+\rm{p}+\rm{th}}$ as more of the thermal energy has been converted into kinetic energy.

Therefore, like the final separation, the amount of unbound mass increases with the initial internal energy content of the envelope. Particularly, this means it is important to include radiation pressure when modelling a massive star \ac{CE}, without which the fraction of unbound mass would be significantly underestimated, by up to 56 per cent in the present case. This difference is expected to be even larger for more massive donors, as they are more radiation-dominated. Moreover, our results suggest that recombination energy can aid envelope ejection, with the full \ac{EoS} simulations that are run for longer (but performed at lower resolution) unbinding the entire envelope. This agrees with similar comparisons made in the regime of low-mass \ac{CE} donors \citep{Reichardt+20}. However, recombination energy may be substantially less efficient in ejecting the envelope if we include radiation transport, which is not modelled in most \ac{CE} hydrodynamical simulations (see Section \ref{subsec:recombination}).
\subsection{Energy and angular momentum conservation}
\label{sec:conservation}
Our highest-resolution simulations conserve total energy to within 0.05 per cent and angular momentum to within 0.02 per cent. This high degree of conservation is possible due to using a single, global time-step for all \ac{SPH} particles, although this limits the number of particles we can feasibly use.

\begin{figure}
	\centering
	\includegraphics[width=\linewidth]{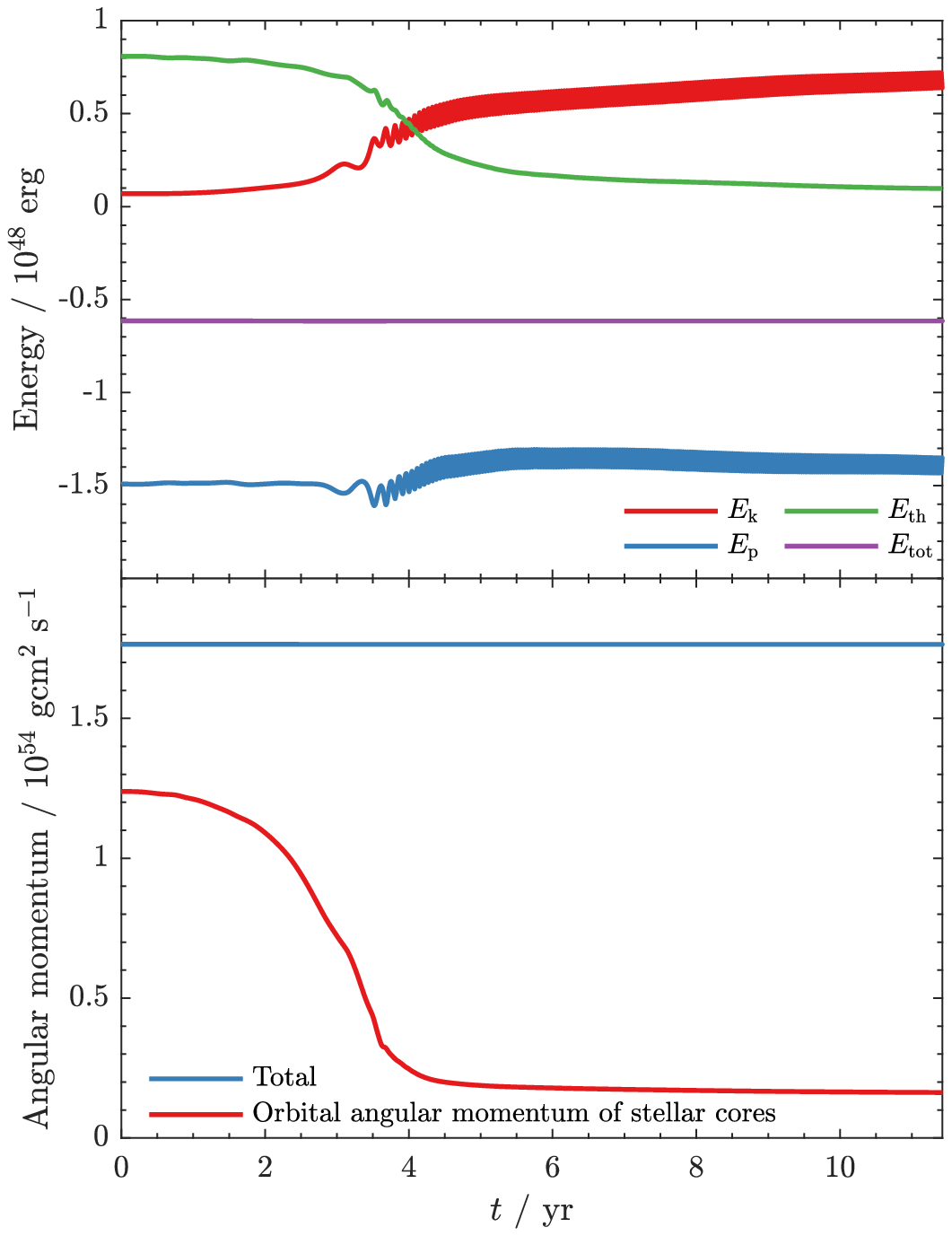}
	\caption{Energy and angular momentum conservation in the high-resolution (2M particles) $q=0.25$ simulation assuming the gas + radiation equation of state. \textit{Top panel:} Evolution in total kinetic energy, $E_\mathrm{k}$ (red), total potential energy $E_\mathrm{p}$ (blue), total thermal energy, $E_\mathrm{th}$ (green), and total energy, $E_\mathrm{tot} = E_\mathrm{k} + E_\mathrm{p} + E_\mathrm{th}$ (purple). \textit{Bottom panel:} Evolution in total angular momentum (blue) and the orbital angular momentum of the donor core and companion (red).}
	\label{fig:conservation}
\end{figure}

The top panel of Fig. \ref{fig:conservation} shows the evolution of kinetic energy, $E_\text{k}$, potential energy, $E_\text{p}$, thermal energy, $E_\text{th}$, and total energy, $E_\text{tot}$, while the bottom panel shows the evolution of total angular momentum and the orbital angular momentum of the stellar cores. We use results from the high-resolution $q=0.25$ simulation assuming the gas + radiation \ac{EoS}.

\section{Discussion} \label{sec:discussion}
\subsection{Mass-loss mechanism}
\label{subsec:unbinding_mech}
\begin{figure}
	\centering
	\includegraphics[width=\linewidth]{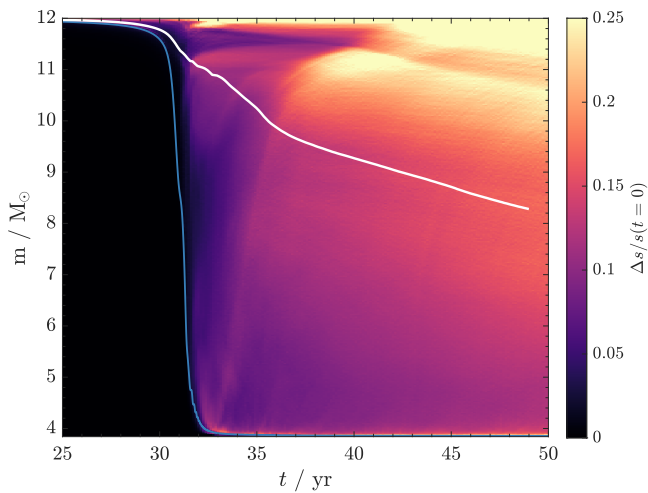}
	\caption{Spherically-averaged entropy profile over the course of the medium-resolution $q=0.25$ simulation assuming the gas + radiation equation of state. The colour indicates the fractional gain in specific entropy, $\Delta s/s(t=0)$, relative to the initial entropy. The blue line gives the mass coordinate of the companion relative to the donor core position with a minimum value of $m_\text{core} = 3.84$\Msun, while the white line shows the amount of bound mass (as defined in Section \ref{subsec:unbinding}).}
	\label{fig:entropy_1dprof}
\end{figure}
\begin{figure*}
	\centering
	\includegraphics[width=0.72\linewidth]{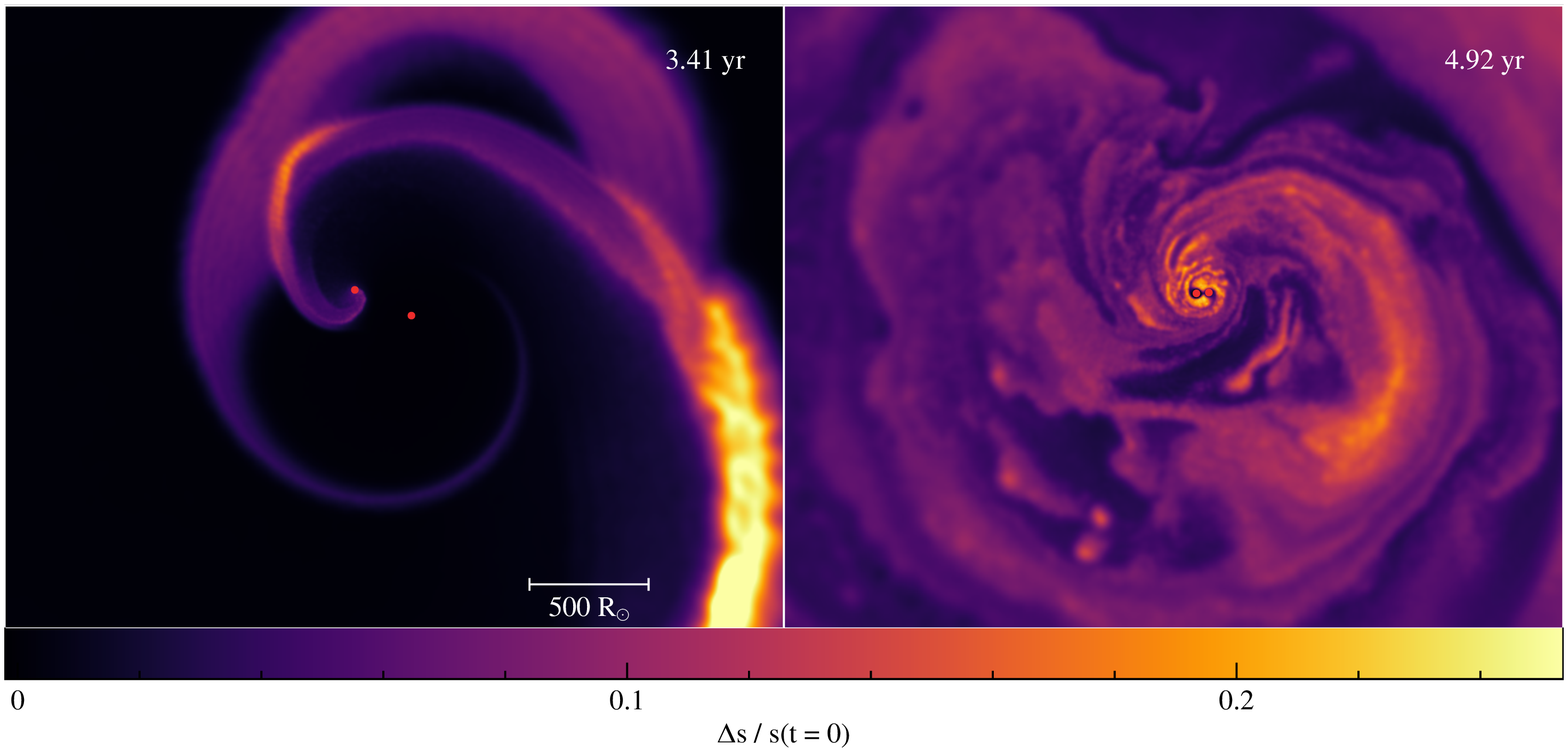}
	\caption{Fractional gain in specific entropy relative to the initial entropy, $\Delta s/s(t=0)$, in the equatorial ($z = 0$) cross-section. \textit{Left panel}: Snapshot during the early plunge-in ($t=3.41$ yr) of the $q=0.25$ gas + radiation simulation at high resolution. \textit{Right panel}: Snapshot during the slow spiral-in ($t=4.92$ yr) of the same simulation.}
	\label{fig:entropy_cross_sec}
\end{figure*}

In this section, we describe the stages of envelope gas becoming unbound in our simulations and identify energy injection mechanisms in the \ac{CE}. Comparison of the top and bottom panels of Fig. \ref{fig:unbound_mass} shows that unbinding occurs in three stages: (i) Unstable \ac{RLOF} preceding the plunge-in, (ii) Dynamical plunge-in, and (iii) Slow spiral-in. In the first phase ($t-t_0 \lesssim 0.2$ yr), 4-5 per cent of the donor envelope becomes unbound via the $L_2$ point. The low- and medium-resolution simulations appear to eject more material than the high-resolution simulations because of their larger initial separations, which lead to longer \ac{RLOF} episodes. 

In the second phase, the plunge-in, energy injection occurs due to shock heating from the companion's bow shock and direct expulsion from the companion's orbital motion. This is seen as the steepening of the unbound mass curve from around $t-t_0\approx0$--2 yr. Around 5--15 per cent of the envelope mass becomes unbound over a few years depending on the \ac{EoS}, increasing from the ideal gas \ac{EoS} to the gas + radiation and full \ac{EoS}. 

More of the envelope becomes gradually unbound during the subsequent slow spiral-in ($t-t_0\gtrsim 2$ yr), which lasts decades after the plunge-in and has the most noticeable \ac{EoS} dependence. Fig. \ref{fig:unbound_mass} shows that the full \ac{EoS} simulation results in the most unbound mass, followed by the gas + radiation and ideal gas \ac{EoS} simulations. The fraction of unbound mass during the slow spiral-in is also the most resolution-dependent.

To examine the sources of energy injection in each phase, we plot in Fig. \ref{fig:entropy_1dprof} the spherically-averaged fractional gain in specific entropy, $\Delta s/s(t=0)$, since the start of the simulation. The expression for $s$ is given in equation (\ref{eq:entropy}). While the plot shows the simulation assuming the gas + radiation \ac{EoS}, the ideal gas and full \ac{EoS} cases display qualitatively similar behaviour. The blue curve shows the mass coordinate of the companion relative to the donor core, while the white curve shows the amount of bound envelope mass. We observe that the entropy profile changes from completely black to purple as the companion plunges in ($t\approx$ 29--33 yr). Because there is no energy transport in our simulations, the only way of generating entropy is through shocks. In this case, the high-entropy material originates from the spiral wake of the companion. This can be seen in Fig. \ref{fig:entropy_cross_sec}, which shows $\Delta s/s(t=0)$ in the $z=0$ cross-section during plunge-in. The left panel displays two shocks. The main shock is created by the companion's supersonic motion through the envelope. Further shock heating occurs where the companion's curved Mach cone overlaps with the wake produced during the previous orbit (bright orange-yellow regions), increasing the specific entropy to $\gtrsim 1.2$ times its initial value. The other shock is created by tidal distortion of the donor's surface as the companion is about to enter the envelope. Velocity perturbations in the rest of the donor star are subsonic, hence do not lead to entropy generation and appear black in the Figure.

Fig. \ref{fig:entropy_1dprof} shows continuous energy injection during the slow spiral-in by the tightly orbiting stellar cores at the centre of the \ac{CE}. Energy injection occurs near $m=4$\Msun, and heats the bulk of the envelope gradually, as indicated by the steadily brightening plot colour after $\approx 33$ yr. Over the same period, the amount of bound mass declines by $\approx 30$ per cent. The right panel of Fig. \ref{fig:entropy_cross_sec} shows entropy production in the vicinity of the stellar cores, consistent with Fig. \ref{fig:entropy_1dprof}. We find qualitatively similar results for the ideal gas \ac{EoS} case, but with smaller entropy increases and less ejected material.

\subsubsection{Role of recombination}
\label{subsec:recombination}

\begin{figure}
	\centering
	\includegraphics[width=\linewidth]{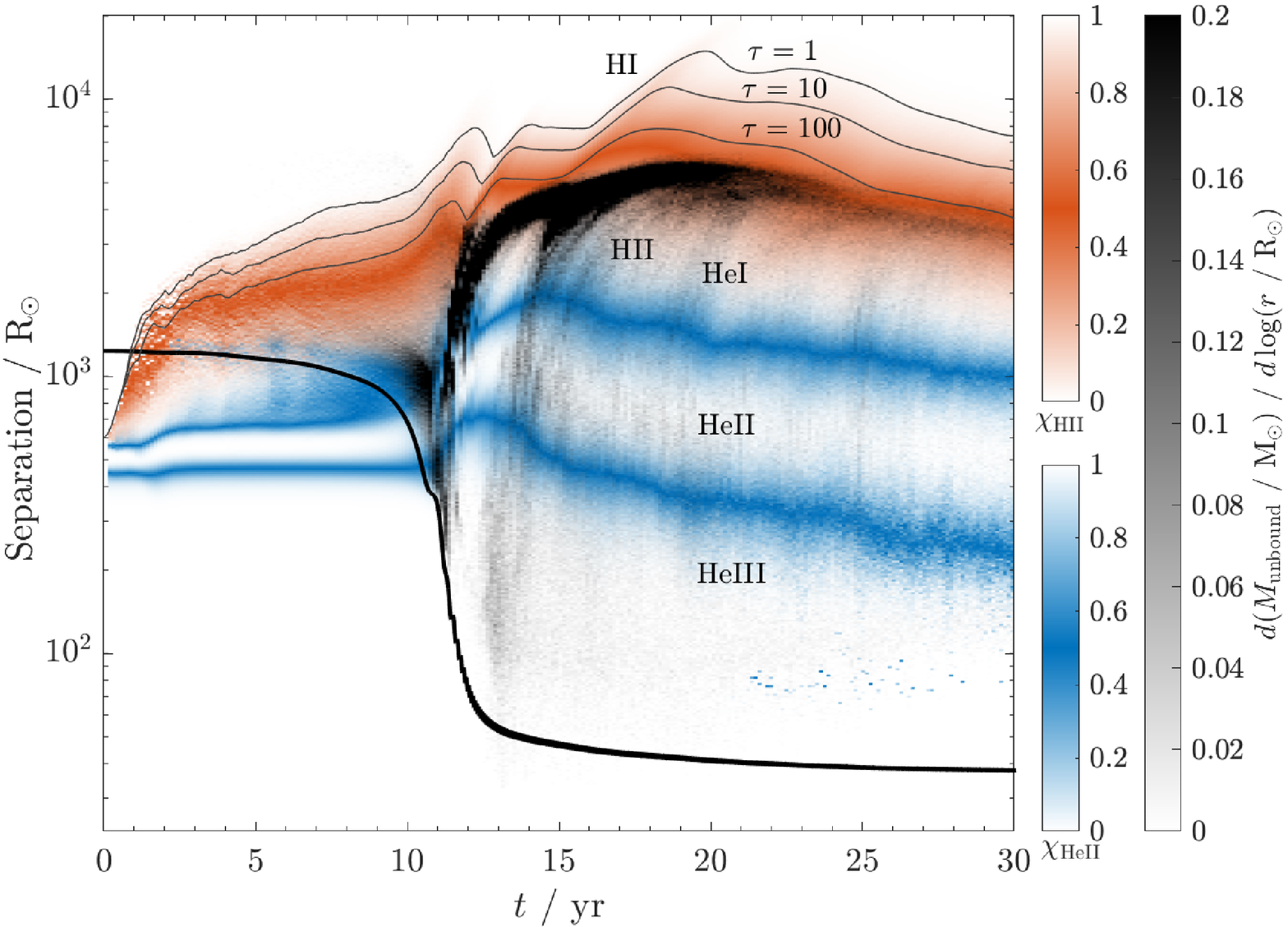}
	\caption{The coloured regions show the spherically-averaged locations at which the envelope is actively recombining over the course of the $q=0.25$ full \ac{EoS} simulation at medium resolution. The \HI, \HeI, and \HeII partial-ionisation zones are shown as the red, upper blue, and lower blue shaded regions, respectively, where the darkest shading occurs at an ionisation fraction of $\chi=0.5$. The black shaded regions indicate the time and radius at which envelope material becomes unbound. Also plotted is the core-companion separation (thick black curve) and contour lines of spherically-averaged optical depth at $\tau = 1,10,100$ (thin black curves).}
	\label{fig:ion_profile}
\end{figure}

In the full \ac{EoS} simulation, there is additional energy injection from hydrogen and helium recombination. While accurately modelling the deposition of recombination energy in the \ac{CE} requires radiation transport, our adiabatic simulations can still provide upper limits on the efficiency of recombination energy.

Fig. \ref{fig:ion_profile} shows the spherically-averaged regions where recombination takes place over the course of the full \ac{EoS}, $q=0.25$ simulation at medium resolution. The red and blue shaded regions represent the locations where hydrogen and helium are actively recombining, respectively. The black patches show the spatial distribution of gas that acquires positive energy, and so is marked unbound, over the course of the simulation.

The black patches appear to trace the bottom of the partially-ionised hydrogen region during the slow spiral-in ($t \gtrsim 12.5$ yr), at $\approx 4,000-5,000$\Rsun. This is well above the partially-ionised \HeI and \HeII regions, meaning almost all the unbound material gains the full helium recombination energy prior to becoming unbound. For our donor, this amounts to $9.9\times10^{46}$ erg in the entire envelope, which is 19 per cent of the envelope binding energy (including thermal energy, see Fig. \ref{fig:ebind_profile}). On the other hand, hydrogen recombination is less important energetically, as $\lesssim 40$ per cent of hydrogen has recombined prior to unbinding. This suggests that hydrogen recombination energy can be regarded as a source of inefficiency in \ac{CE} evolution, as it mainly supplies energy to material after it has been unbound (see discussion in Section \ref{subsec:alpha}).

Because our simulations are adiabatic, recombination energy cannot be radiated away. We check the validity of this assumption by calculating the optical depth of the recombining regions. We integrate the spherically-averaged optical depth, $\tau$, radially inwards from infinity, using post-processed opacities calculated with the \MESA opacity tables as implemented in \Phantom \citep{Reichardt+20}. Fig. \ref{fig:ion_profile} shows three contour lines corresponding to $\tau=1,10,100$ as labelled in the Figure. \HeIII and \HeII recombination takes place in optically-thick regions, where the recombination energy is not expected to be transported away by radiative flux. This is less clear for hydrogen recombination, since the photosphere, which approximately follows the $\tau=1$ contour, traces the upper edge of the hydrogen partial ionisation zone. However, as mentioned previously, $\lesssim 40$ per cent of hydrogen has recombined prior to unbinding, and Fig. \ref{fig:ion_profile} shows that this layer can still be optically-thick ($\tau>100$). 

Our findings are broadly consistent with studies of low-mass star \ac{CE} simulations, which find that while helium recombination energy is likely to be used for ejecting the envelope, hydrogen recombination energy is less energetically useful \citep{Ivanova+13b,Ivanova+15,Reichardt+20}. Moreover, the efficiency of hydrogen recombination has been debated in the literature \citep{Sabach+17,Grichener+18,Ivanova18,Soker+18}. These discussions focus on the possibility for hydrogen recombination energy to be transported to the surface by convection or radiation, where it will be radiated away without doing work on the envelope. Our simulations do not fully model this process, as they do not include radiation transport and adequately model convection (see Section \ref{subsec:energy_transport}). Moreover, if convection rather than radiation is the dominant form of energy transport in the hydrogen partial ionisation zone, our optical depth analysis is irrelevant, and one should instead estimate whether the convective flux is sufficiently large to transport away recombination energy \citep[e.g.][]{Soker+18}.

It is possible that the initial convective flux in a \ac{RSG} envelope, which we do not model, transports part of the helium recombination energy away to the surface before it can drive envelope expansion. In our simulations, the total helium recombination energy ($10^{47}$ erg) is released over $\approx 10$ yr starting from the plunge-in, implying an average luminosity of $10^{47} ~\mathrm{erg}/10~\mathrm{yr} = 3\times10^{38}~\mathrm{erg}~\mathrm{s}{}^{-1}$. This is larger than the \ac{RSG}'s initial luminosity, $1.7\times10^{38}~\mathrm{erg}~\mathrm{s}{}^{-1}$, and so it is unlikely for this energy to be efficiently carried out by the initial star's energy flux alone.

Nonetheless, we have shown that (i) helium recombination dominates energetically in our simulations, and is deposited much deeper inside the \ac{CE}, where it is unlikely to be transported away, and (ii) material is marked unbound in regions with rather high \HII fractions ($\gtrsim 60$ per cent). Therefore, the much larger amount of unbound mass observed in our full \ac{EoS} simulations compared to the gas + radiation \ac{EoS} simulations depends little on the ability for hydrogen recombination energy to do work on the envelope.

\subsection{Termination of the dynamical plunge-in}
\label{subsec:termination}

We explore the conditions and uncertainties surrounding the termination of the dynamical plunge-in observed in our simulations, and discuss possible interactions that may further shrink the binary orbit beyond what we have simulated.

We find that at the end of the plunge-in, the stellar cores have evacuated most of the gas in the vicinity of their orbit and spun up the small amount of remaining gas. Both of these effects significantly decrease the gravitational drag the stellar cores experience. The blue curve in Fig. \ref{fig:entropy_1dprof} shows the mass coordinate of the companion in the $q=0.25$ gas + radiation \ac{EoS} simulation, revealing that very little mass remains inside the orbit after the plunge-in. For instance, at $-P_\mathrm{orb}\dot{a}/a = 5\times10^{-4}$ ($t=38.8$ yr), there is only 0.02\Msun of material between the companion and the $m_\mathrm{core}=3.84\Msun$ core of the donor. This material is concentrated near the stellar cores and co-rotating with the orbit.

\begin{figure}
	\centering
	\includegraphics[width=\linewidth]{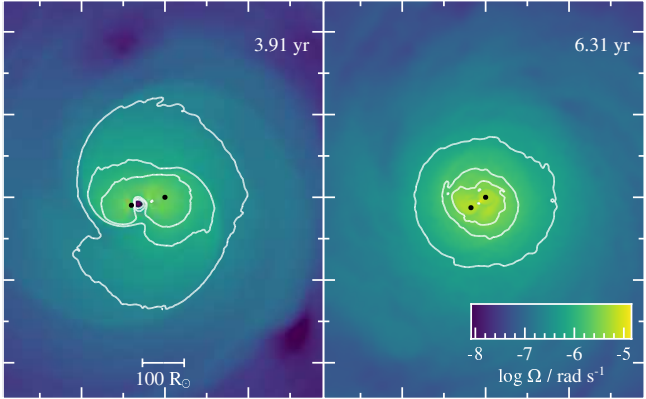}
	\caption{$z$-component of the gas angular velocity, $\Omega$, in the orbital plane ($z=0$) at $t =$ 3.91 yr (left panel) and 6.31 yr (right panel) for the $q=0.25$, gas + radiation equation of state simulation at high resolution. From outside towards the centre, the three white contours are surfaces of $\Omega/\Omega_\text{orb} = 0.1, 0.3, 0.5$, which is the ratio of the gas angular velocity to the Keplerian angular velocity.}
	\label{fig:omega_xy}
\end{figure}

Fig. \ref{fig:omega_xy} shows the $z$-angular velocity, $\Omega$, in the orbital plane for the $q=0.25$ gas + radiation \ac{EoS} simulation at high resolution. $\Omega$ is calculated relative to the stellar cores' centre of mass. We display two snapshots, taken at the steepest point of the plunge-in (left panel) and near the beginning of the slow spiral-in (right panel). To measure the degree of co-rotation of the gas, we include three contour lines of $\Omega/\Omega_\text{orb}$, which is the ratio of the gas angular velocity to the orbital angular velocity of the stellar cores. This ratio increases from the outermost to the innermost contour line ($\Omega/\Omega_\text{orb} = 0.1,0.3,0.5$). In the left panel, where the companion is rapidly plunging, only gas that is tightly bound to the stellar cores is in co-rotation, as seen from the fact that the innermost contour wraps around the individual stellar cores. In the right panel, which depicts the end of the dynamical inspiral, the same contours become circular, indicating that gas along the orbital path uniformly rotates with the stellar cores. Our results agree with \cite{Reichardt+2019}, whose low-mass \ac{CE} simulations find that the gas co-rotates with the orbit within few tens of percent by the end of the plunge-in. On the other hand, \cite{Ivanova&Nandez16} find in their simulations that the orbital angular velocity of the stellar cores exceeds the angular velocity of the surrounding material by about an order of magnitude even after orbital stabilisation.

\begin{figure*}
	\centering
	\includegraphics[width=0.85\linewidth]{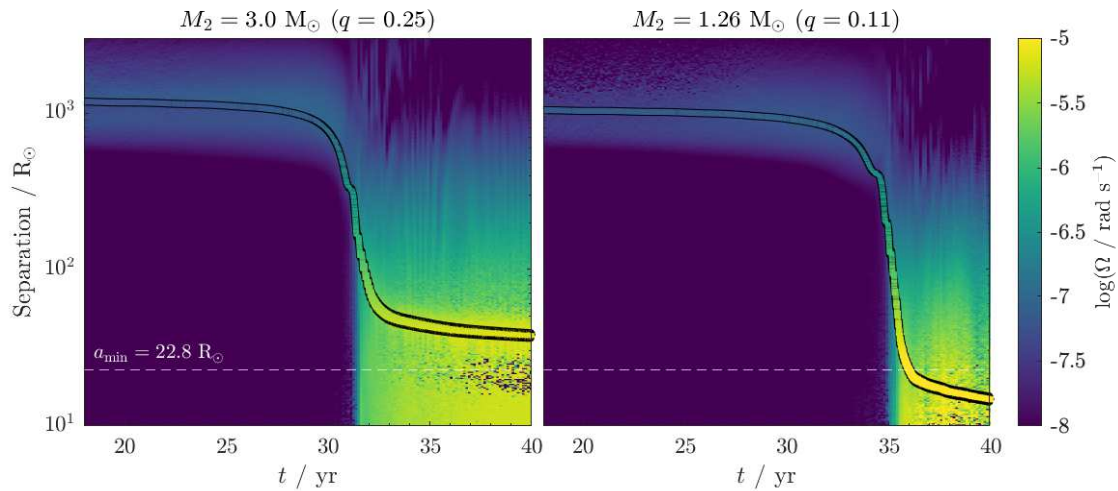}
	\caption{Evolution in the cylindrically-averaged $z$-angular velocity of the envelope gas, $\Omega$, over the course of the medium-resolution gas + radiation \ac{EoS} simulation. The left panel shows the $q=0.25$ simulation while the right panel shows the $q=0.11$ simulation. We also plot the core-companion separation, whose colour shows the orbital angular velocity of the stellar cores.}
	\label{fig:omega_profile}
\end{figure*}

Fig. \ref{fig:omega_profile} plots the evolution in the cylindrically-averaged $\Omega$ relative to the donor core. We also plot the core-companion separation, colouring the curve to indicate the instantaneous orbital angular velocity of the stellar cores. The colour difference between the separation curve and the background therefore shows the velocity contrast between the stellar cores and the surrounding gas. Prior to the plunge-in, the dark purple $\Omega$-profile reflects the initially non-rotating donor star, with only a small amount of surface material having been torqued by the companion. During the plunge-in, the \ac{CE} spins up to a few times $10^{-6}$ rad s${}^{-1}$. For the $q=0.25$ simulation (left panel), the stellar cores spin up the gas in their vicinity for $t\gtrsim 34$ yr, as seen from the closeness in colour between the inspiral curve and the surrounding matter. This may be contrasted with the $q=0.11$ simulation (right panel), where the companion continues to spiral beneath the minimum separation of 22.8\Rsun. The stellar cores rotate faster than their surrounding gas, as the colour of the separation curve is brighter than its surroundings.

Our simulations show that the end of the plunge-in is therefore coincident with reduced gas density near the stellar cores and reduced velocity contrast between the stellar cores and this remaining gas. However, there are a number of effects that can drive the orbit to tighten further. Firstly, we have not demonstrated complete envelope ejection in our simulations (for the full \ac{EoS} case, this has only been demonstrated at low- and medium-resolution). This means that the expanded but still gravitationally bound gas could fall back and drive further spiral-in. Secondly, the near absence of gas remaining inside the orbit suggests that there could be more gas evacuated from deeper unresolved layers that are part of the numerical core. However, because these unresolved layers are radiative, they should expand on a longer thermal time-scale and lead to stable mass transfer instead of resuming a dynamical inspiral. \citet{Vigna-Gomez+21} carried out a 1D study using a less evolved 12\Msun \ac{RSG} model. They found that stripping to the lower convective boundary may lead to re-expansion on $\sim 10^3$ yr time-scales to few tens of solar radii, though the maximum radius reached during expansion decreases as a larger amount of the envelope is stripped/ejected. 

\subsection{Common-envelope efficiency} \label{subsec:alpha}
Using the final orbital separations derived from our simulations, we estimate upper limits to the \ac{CE} efficiency parameter $\alpha$ \citep{Webbink1984,Livio&Soker88}, defined through the equation
\begin{align}
	-\alpha\Delta E_\mathrm{orb} = E_\mathrm{bind},
	\label{eq:alpha-lambda}
\end{align}
where $E_\mathrm{bind} > 0$ is the envelope binding energy and $\Delta E_\mathrm{orb} < 0$ is the difference between the final and initial orbital energy,
\begin{align}
	\Delta E_\mathrm{orb} =
	- \frac{Gm_\mathrm{1,core} M_2}{2a_f}
	+ \frac{GM_1 M_2}{2a_i},
	\label{eq:Eorb}
\end{align}
where $a_i$ and $a_f$ are the initial and final semi-major axes, $M_2$ is the accretor mass, $M_1 = 12\Msun$ is the donor mass, and $m_\mathrm{1,core}$ is the donor's core mass, represented in our simulations by the 3.84\Msun point mass. Equation (\ref{eq:alpha-lambda}) is based on global energy conservation, asserting that the energy used for unbinding the envelope must be ultimately sourced from the orbital energy released by the inspiralling stellar cores, with some efficiency $\alpha$.

Our calculations only provide upper limits on $\alpha$, because (i) we do not eject (do not conclusively eject in the case of the full \ac{EoS}) the entire envelope in every simulation, and (ii) the orbital separation continues to decline slowly at the end of our simulations, and so the final orbital separation is an upper limit to $a_f$. As a result, the values of $\alpha$ we present are not intended for determining the outcome of \ac{CE} events. Nonetheless, it is informative to compare upper limits on $\alpha$ calculated with three different definitions of $E_\mathrm{bind}$, which allows us to quantify the ability for thermal and recombination energy to do work on the envelope:
\begin{enumerate}
	\item $\alpha_\mathrm{grav}$: We take $E_\mathrm{bind}$ to be the gravitational potential energy of the envelope, which is independent of the assumed \ac{EoS} (following Section \ref{subsec:eos}).
	\item $\alpha_\mathrm{th}$: We include the contribution of thermal energy (not including recombination energy) in $E_\mathrm{bind}$. This reduces the magnitude of the total binding energy, which implies $\alpha_\mathrm{th} < \alpha_\mathrm{grav}$ always.
	\item $\alpha_\mathrm{int}$: We include the full internal energy in $E_\mathrm{bind}$, as in equation (\ref{eq:ebind}). This results in a smaller binding energy when assuming the full \ac{EoS} due to the presence of recombination energy, which implies $\alpha_\mathrm{int} < \alpha_\mathrm{th}$. With the ideal gas and gas + radiation \acp{EoS}, $\alpha_\mathrm{int} = \alpha_\mathrm{th}$.
\end{enumerate}

Table \ref{tab:alpha} displays the three $\alpha$ parameters calculated for each simulation. We do not include results for the $q=0.11$ simulations, as the companion fails to stall within the simulation. For the $q=0.25$ simulations, $\alpha_\mathrm{grav}$ increases from the ideal gas \ac{EoS} to the gas + radiation and full \ac{EoS} simulations due to their increasing final separations. The values of $\alpha_\mathrm{grav}$ are all substantially above unity, mainly because the envelope has not yet been fully unbound and because internal energy has not been accounted for. Accounting for thermal energy, the values for $\alpha_\mathrm{th}$ are much closer to unity and also display a smaller spread across the simulations with different \acp{EoS}. While $\alpha_\mathrm{th}$ is similar for the ideal gas and gas + radiation \acp{EoS}, it is 20 per cent larger for the full \ac{EoS} case due to additional energy from recombination that is not included in $E_\mathrm{bind}$. When accounting for this source, we obtain a much smaller value, $\alpha_\mathrm{int} = 0.58$. Although the full \ac{EoS} simulations eject a much larger fraction of the envelope, our analysis in Section \ref{subsec:recombination} shows that a very small fraction of hydrogen recombination energy is useful.

\begin{table}
	\centering
	\begin{tabular}{llll}
		\toprule
		EoS & $\alpha_\text{grav}$ & $\alpha_\text{th}$ & $\alpha_\text{int}$ \\ \midrule
	    Ideal gas  &  2.0  & 1.0   & 1.0   \\
		Gas + rad. &  2.3  & 0.97  & 0.97  \\
		Full       &  2.8  & 1.2   & 0.58  \\
		\bottomrule
	\end{tabular}
	\caption{Values of the common-envelope efficiency parameter, $\alpha$, inferred from the final separation, but with various definitions of the binding energy (see Section \ref{subsec:alpha}). We only report values for the $q=0.25$ simulations, as the companion fails to stall in the $q=0.11$ simulations.}
	\label{tab:alpha}
\end{table}

However, $\alpha_\mathrm{th}$ and $\alpha_\mathrm{int}$, as defined, implicitly assume that thermal and ionisation energy, respectively, are fully efficient energy sources. This is because the efficiency parameter $\alpha$ only multiplies the orbital energy term. Here, we assume that there are distinct efficiencies, $\tilde{\alpha}_\mathrm{orb}$, $\tilde{\alpha}_\mathrm{th}$, and $\tilde{\alpha}_\mathrm{rec}$, for the orbital, initial thermal, and recombination energy, respectively. Assuming that each efficiency is universal among the simulations, this formalism can be applied to each of our simulations with different \acp{EoS} to give the following equations:
\begin{align}
	-\tilde{\alpha}_\text{orb}\Delta E_\text{orb}^\text{ideal} + \tilde{\alpha}_\text{th} E_\text{th}^\text{ideal} &= -E_\text{grav}, \label{eq:tilde_alpha_a} \\
	-\tilde{\alpha}_\text{orb}\Delta E_\text{orb}^\text{gasrad} + \tilde{\alpha}_\text{th} E_\text{th}^\text{gasrad} &= -E_\text{grav}, \label{eq:tilde_alpha_b} \\
	-\tilde{\alpha}_\text{orb}\Delta E_\text{orb}^\text{full} + \tilde{\alpha}_\text{th} E_\text{th}^\text{full} + \tilde{\alpha}_\text{rec} E_\text{rec}^\text{full} &= -E_\text{grav}, \label{eq:tilde_alpha_c}
\end{align}
where the superscript indicates the \ac{EoS} corresponding to the quantity, while the subscript refers to the component of the envelope energy, with $E_\text{th}$ being the initial thermal energy, $E_\text{rec}$ being recombination energy, and $E_\text{grav} < 0$ being the gravitational potential energy. Solving equations (\ref{eq:tilde_alpha_a})-(\ref{eq:tilde_alpha_c}) yields $\tilde{\alpha}_\text{orb} = 1.2$, $\tilde{\alpha}_\text{th} = 0.81$, and $\tilde{\alpha}_\text{rec} = 0.43$. This allows us to directly infer that the initial thermal energy is only around 68 per cent as efficient as orbital energy, but 1.9 times the efficiency of recombination energy. Despite its low efficiency, recombination energy is still able to significantly aid envelope ejection in our simulations, as it comprises a large fraction of the total energy in the envelope we defined.

The current analysis could have important implications for rapid population synthesis, where one wishes to determine the final separation of a \ac{CE} event using equation (\ref{eq:alpha-lambda}) without knowing the detailed stellar structure of the donor. Thus, the binding energy is usually expressed as $E_\mathrm{bind} = GM_1(M_1-m_{1,\mathrm{core}})/(\lambda R_1)$ via a parameter $\lambda$ \citep{deKool1990} that is tabulated or analytically calculable from gross properties of the donor such as the evolutionary stage, mass, and radius \citep[e.g.][]{Xu&Li2010,Loveridge+11,Kruckow+16}. Values of $\lambda$ that include or exclude the envelope internal energy have been explored by the literature. Our analysis results tentatively support the use of $\lambda$ values that include the contribution of envelope thermal energy, but not the full recombination energy. Further study including radiation transport is needed to confirm this.
\subsection{On our adiabatic treatment}
\label{subsec:energy_transport}
As discussed in Section \ref{subsec:recombination}, the adiabatic nature of our simulations implies that recombination energy is fully thermalised in the envelope. On the other hand, the retention of the recombination energy in the envelope, particularly hydrogen recombination energy, is unclear \citep{Sabach+17,Grichener+18,Soker+18}, with our analysis confirming that part of hydrogen recombination occurs in optically-thin layers, although mainly in material that has already been unbound. A full simulation including radiation transport is needed to assess how much recombination energy should instead be radiated away.

The adiabatic assumption is more generally problematic for massive stellar donors, as the envelope thermal time-scale may be shorter than or comparable with the dynamical time-scale \citep{Ivanova+2013,Ricker+2018}. Taking our \MESA 12\Msun \ac{RSG} reference model as an example, the integrated thermal time of the convective envelope is $\tau_\mathrm{th} = \int_{m_\mathrm{core}}^M c_P T dm / L_\mathrm{surf} = 480$ yr, where $c_P$ is the specific heat capacity at constant pressure, and $L_\mathrm{surf}$ is the surface luminosity. Furthermore, we have not modelled the significant amount of energy injected into the \ac{CE} due to nuclear burning, which is likely to impact the late slow spiral-in. For the unperturbed donor star, the amount of energy flowing in and out of the envelope over a duration $\Delta t = 40~\mathrm{yr}$ at the stellar luminosity $L = 4.4\times 10^4\Lsun$ is $L\Delta t = 2.2\times 10^{47}~\mathrm{erg}$, which is comparable to the envelope binding energy and initially available recombination energy.

A closely related issue is the need for modelling convection in a \ac{CE}, as it may have a significant impact on the ability for orbital and recombination energy to do work on the envelope \citep{Sabach+17,Grichener+18,Soker+18,Wilson&Nordhaus19,Wilson&Nordhaus20}. However, this is not possible with an adiabatic simulation, as convection requires a negative entropy gradient to be sustained by central nuclear burning and surface cooling. Furthermore, \cite{Ricker+2018} has made preliminary comparisons between a \ac{CE} simulation with and without radiation transport, finding that the simulation without radiation transport grossly overestimates convective velocities, as radiation should carry a significant portion of the energy flux.

We also emphasise that modelling convection is required for creating stable, 3D hydrodynamical models of \acp{RSG}. In experiments we performed, we found that the initial entropy gradient in the convective envelope is erased as high-/low-entropy material rises/sinks without turning over, resulting in a transient surface expansion. We thus circumvented this issue by setting up an artificial envelope profile that suppresses convection by fixing the entropy profile to be constant relative to the gas + radiation \ac{EoS} (see Section \ref{subsec:donor}). However, when using the full \ac{EoS}, the same density/pressure profile still gives rise to a negative entropy gradient, as the recombined surface material has a lower entropy. This results in an initial surface expansion for the full \ac{EoS} case (Fig. \ref{fig:stability}c), as was also observed by \cite{Ohlmann+2017}. The main effect of this surface expansion is to reduce the time between the start of the simulation and the plunge-in. We do not expect it to significantly alter the dynamics of the plunge-in and the final outcome of the \ac{CE} interaction, which are mainly determined by the less tenuous layers below.

Thus, to shed light on the unclear role of recombination energy, or even just to model a stable and convecting giant envelope, future efforts should be directed towards using radiation hydrodynamics. However, our present study using fully adiabatic simulations still allows for a controlled experiment that determines the impact of the initial internal energy content on \ac{CE} evolution.
\subsection{Astrophysical implications}
\label{subsec:implications}
This study has explored the outcome of a \ac{CE} phase involving a 12\Msun donor with a 1.26\Msun ($q=0.11$) and with a 3\Msun ($q=0.25$) companion. The 1.26\Msun companion could be interpreted as a neutron star, in which case our simulations represent the \ac{CE} phase experienced by progenitors of Galactic double neutron stars and merging double neutron stars \citep{Bhattacharya+vandenHeuvel91,Tutukov&Yungelson93,Belczynski+02,Tauris+2017,Kruckow+18,Vigna-Gomez+2018,Vigna-Gomez+20}. However, we have not resolved any potential feedback from the neutron star, as the companion softening length, $h_2=2.15\Rsun$, is much larger than the $\approx 11$ km neutron star radius. A bigger issue is that our $M_2=1.26\Msun$ simulations have not determined the fate of the post-\ac{CE} system (see Section \ref{subsec:sep}). The companion may stall beneath the convective part of the envelope \citep{Fragos+2019,Law-Smith+20}. If the neutron star fails to stall, it may merge with the helium core, leading to a merger-induced explosion or `\ac{CE} jets supernova' \citep{Soker&Gilkis18,Gilkis+19,Grichener&Soker19,Soker+19,Schroder+20,Dong+21} or form a Thorne-\.Zytkow object \citep{Thorne+Zytkow77,Podsiadlowski+95}.

If the neutron star does stall, the remaining hydrogen layers bound to the donor core are expected to re-expand on a thermal time scale \citep{Vigna-Gomez+21}, possibly leading to another mass transfer episode. This phase was explored in the 1D \ac{CE} simulation by \cite{Fragos+2019}, who also modelled a 12\Msun donor star. They found post-\ac{CE} separations of 10\Rsun, and predicted that continued evolution during the self-regulated inspiral and subsequent stable mass transfer episode will further tighten the binary to a few solar radii, allowing it to form a double neutron star that will merge within a Hubble time. Thus, to make progress, future simulations should explore accretion/feedback boundary conditions for the neutron star companion, and resolve deeper layers of the \ac{RSG} envelope. \cite{Law-Smith+20} made recent progress in the latter, having fully resolved the helium core by restricting their simulation to the last 8\Rsun ($\sim 1$ per cent) of the dynamical inspiral.

Our simulations with a 3\Msun companion could represent the \ac{CE} episode experienced by less massive ($\lesssim 20\Msun$) progenitors of stripped-envelope \acp{SN} \citep{Podsiadlowski+92,Eldridge+08,Eldridge+15,Yoon+10,Yoon15,Zapartas+19,Sravan+19}, where binary interactions are likely to have removed most or all of the hydrogen envelope. In the former case, the \ac{SN} would be classified as type IIb, whereas in the latter case, the \ac{SN} would be classified as type Ib. Since the numerical core in our simulation represents the helium core and also part of the tightly bound hydrogen envelope, it is expected to become a type IIb \ac{SN} progenitor unless there is further mass-loss via mass transfer or winds after the CE phase.

\ac{CE} interaction is also important for forming the progenitors of core-collapse \acp{SN} that exhibit ejecta-companion interaction, where the \ac{SN} ejecta interact with a companion star in a (most likely) post-\ac{CE} binary system \citep{Hirai+14,Hirai&Yamada15,Liu+15,Rimoldi+16,Hirai+18,Hirai+2020}. In particular, SN 2006jc, a type Ibn \ac{SN} with an observed companion, has recently been proposed to be a possible stripped-envelope \ac{SN} in which a main-sequence companion has been inflated by ejecta-companion interaction \citep{Sun+20,Ogata+21}. Estimates of the SN 2006jc host environment age by \cite{Sun+20} are consistent with the lifetime of a 12-15\Msun star. Moreover, the ejecta-companion interaction models by \cite{Ogata+21} constrain the mass of the main-sequence companion to be $\sim 3\Msun$ and the pre-\ac{SN} orbital separation to be $\sim 40\Rsun$. Our $q=0.25$ simulations have parameter values that are consistent with such a progenitor, and find post-\ac{CE} separations that are consistent with their model estimates.

\section{Summary and conclusions} \label{sec:conclusion}
We performed 3D, global hydrodynamical simulations of a \ac{CE} involving a massive star donor. We used a 12\Msun \ac{RSG} donor with a 3\Msun ($q=0.25$) and with a 1.26\Msun companion ($q=0.11$). We investigated the contribution of thermal and recombination energy to unbinding a massive star envelope by comparing simulations performed with three different \acp{EoS}: (i) Ideal gas, (ii) Gas + radiation, and (iii) A `full' \ac{EoS} that includes recombination. Using different \acp{EoS} results in a different initial internal energy content of the envelope, increasing from (i) to (iii). Our simulations are adiabatic, and so do not include the effects of radiation transport and convection, which may reduce the efficiency of various energy sources including recombination. Our results therefore place upper limits on the contribution of internal energy to ejecting the envelope. Our main findings are as follows:

\begin{enumerate}
	\item Our combined comparisons of the simulations assuming the ideal gas \ac{EoS} (gas thermal energy), gas + radiation \ac{EoS} (gas and radiation thermal energy), and full \ac{EoS} (gas and radiation thermal energy plus recombination energy) demonstrate that the envelope internal energy could contribute significantly towards ejecting the envelope.
	\item Radiation pressure is generally less significant in low-mass stars, and negligible in the 1-2\Msun donors simulated in many past \ac{CE} studies. For our 12\Msun donor, including radiation pressure doubles the amount of unbound envelope mass (at least 30-69 per cent, depending on the resolution), and increases the final separation by 14 per cent. These differences are expected to be even larger for more massive stars that are more radiation dominated.
	\item Energy injection from recombination further increases the amount of unbound envelope mass by a factor of 2.1, and increases the final separation by an additional 18 per cent.
	\item In the absence of recombination, the unbinding after the dynamical plunge-in is driven by the slow inspiral of stellar cores, which gradually heats the \ac{CE}.
	\item With recombination, the post-plunge-in unbinding is mainly due to helium recombination, which occurs in optically-thick layers of the \ac{CE}. Hydrogen recombination is an inefficient energy source for envelope ejection, as it mostly occurs in already unbound parts of the envelope.
	\item With the $M_2=3.0\Msun$ ($q=0.25$) companion, the companion successfully stalls inside the initial convective envelope of the donor. Whereas with the $M_2=1.26\Msun$ ($q=0.11$) companion, the companion plunges deeper into the radiative intershell that is unresolved by our simulations.
	\item The end of the dynamical plunge-in is associated with significant reduced gas density around the stellar cores and a spin-up of this remaining gas to up to 50 per cent of the orbital angular velocity of the stellar cores.
\end{enumerate}

\section*{Acknowledgements}
M. Y. M. L. acknowledges support by an Australian Government Research Training Program (RTP) Scholarship. IM is a recipient of the Australian Research Council Future Fellowship FT190100574. Parts of this research were supported by the Australian Research Council Centre of Excellence for Gravitational Wave Discovery (OzGrav), through project number CE170100004. Parts of this research work were performed on the Gadi supercomputer of the National Computational Infrastructure (NCI), which is supported by the Australian Government, and on the OzSTAR national facility at the Swinburne University of Technology. The OzSTAR program receives funding in part from the Astronomy National Collaborative Research Infrastructure Strategy (NCRIS) allocation provided by the Australian Government.

\section*{Data availability}
The data underlying this article will be shared on reasonable request to the corresponding author.






\bibliographystyle{mnras}
\bibliography{bibliography.bib}

\begin{thebibliography}{}
\makeatletter
\relax
\def\mn@urlcharsother{\let\do\@makeother \do\$\do\&\do\#\do\^\do\_\do\%\do\~}
\def\mn@doi{\begingroup\mn@urlcharsother \@ifnextchar [ {\mn@doi@}
  {\mn@doi@[]}}
\def\mn@doi@[#1]#2{\def\@tempa{#1}\ifx\@tempa\@empty \href
  {http://dx.doi.org/#2} {doi:#2}\else \href {http://dx.doi.org/#2} {#1}\fi
  \endgroup}
\def\mn@eprint#1#2{\mn@eprint@#1:#2::\@nil}
\def\mn@eprint@arXiv#1{\href {http://arxiv.org/abs/#1} {{\tt arXiv:#1}}}
\def\mn@eprint@dblp#1{\href {http://dblp.uni-trier.de/rec/bibtex/#1.xml}
  {dblp:#1}}
\def\mn@eprint@#1:#2:#3:#4\@nil{\def\@tempa {#1}\def\@tempb {#2}\def\@tempc
  {#3}\ifx \@tempc \@empty \let \@tempc \@tempb \let \@tempb \@tempa \fi \ifx
  \@tempb \@empty \def\@tempb {arXiv}\fi \@ifundefined
  {mn@eprint@\@tempb}{\@tempb:\@tempc}{\expandafter \expandafter \csname
  mn@eprint@\@tempb\endcsname \expandafter{\@tempc}}}

\bibitem[\protect\citeauthoryear{{Belczynski}, {Kalogera}  \&
  {Bulik}}{{Belczynski} et~al.}{2002}]{Belczynski+02}
{Belczynski} K.,  {Kalogera} V.,   {Bulik} T.,  2002, \mn@doi [\apj]
  {10.1086/340304}, \href
  {https://ui.adsabs.harvard.edu/abs/2002ApJ...572..407B} {572, 407}

\bibitem[\protect\citeauthoryear{{Bhattacharya} \& {van den
  Heuvel}}{{Bhattacharya} \& {van den
  Heuvel}}{1991}]{Bhattacharya+vandenHeuvel91}
{Bhattacharya} D.,  {van den Heuvel} E.~P.~J.,  1991, \mn@doi [\physrep]
  {10.1016/0370-1573(91)90064-S}, \href
  {https://ui.adsabs.harvard.edu/abs/1991PhR...203....1B} {203, 1}

\bibitem[\protect\citeauthoryear{{Bodenheimer} \& {Taam}}{{Bodenheimer} \&
  {Taam}}{1984}]{Bodenheimer+Taam84}
{Bodenheimer} P.,  {Taam} R.~E.,  1984, \mn@doi [\apj] {10.1086/162049}, \href
  {https://ui.adsabs.harvard.edu/abs/1984ApJ...280..771B} {280, 771}

\bibitem[\protect\citeauthoryear{{Chamandy} et~al.,}{{Chamandy}
  et~al.}{2018a}]{Chamandy+2018}
{Chamandy} L.,  et~al., 2018a, \mn@doi [\mnras] {10.1093/mnras/sty1950}, \href
  {https://ui.adsabs.harvard.edu/abs/2018MNRAS.480.1898C} {480, 1898}

\bibitem[\protect\citeauthoryear{{Chamandy} et~al.,}{{Chamandy}
  et~al.}{2018b}]{Chamandy+18}
{Chamandy} L.,  et~al., 2018b, \mn@doi [\mnras] {10.1093/mnras/sty1950}, \href
  {https://ui.adsabs.harvard.edu/abs/2018MNRAS.480.1898C} {480, 1898}

\bibitem[\protect\citeauthoryear{{Clayton}, {Podsiadlowski}, {Ivanova}  \&
  {Justham}}{{Clayton} et~al.}{2017}]{Clayton:2017}
{Clayton} M.,  {Podsiadlowski} P.,  {Ivanova} N.,   {Justham} S.,  2017,
  \mn@doi [\mnras] {10.1093/mnras/stx1290}, \href
  {https://ui.adsabs.harvard.edu/abs/2017MNRAS.470.1788C} {470, 1788}

\bibitem[\protect\citeauthoryear{{Cullen} \& {Dehnen}}{{Cullen} \&
  {Dehnen}}{2010}]{Cullen+Dehnen10}
{Cullen} L.,  {Dehnen} W.,  2010, \mn@doi [\mnras]
  {10.1111/j.1365-2966.2010.17158.x}, \href
  {https://ui.adsabs.harvard.edu/abs/2010MNRAS.408..669C} {408, 669}

\bibitem[\protect\citeauthoryear{{De}, {MacLeod}, {Everson}, {Antoni}, {Mandel}
   \& {Ramirez-Ruiz}}{{De} et~al.}{2020}]{De+20}
{De} S.,  {MacLeod} M.,  {Everson} R.~W.,  {Antoni} A.,  {Mandel} I.,
  {Ramirez-Ruiz} E.,  2020, \mn@doi [\apj] {10.3847/1538-4357/ab9ac6}, \href
  {https://ui.adsabs.harvard.edu/abs/2020ApJ...897..130D} {897, 130}

\bibitem[\protect\citeauthoryear{{Diehl}, {Rockefeller}, {Fryer}, {Riethmiller}
   \& {Statler}}{{Diehl} et~al.}{2015}]{Diehl+15}
{Diehl} S.,  {Rockefeller} G.,  {Fryer} C.~L.,  {Riethmiller} D.,   {Statler}
  T.~S.,  2015, \mn@doi [\pasa] {10.1017/pasa.2015.50}, \href
  {https://ui.adsabs.harvard.edu/abs/2015PASA...32...48D} {32, e048}

\bibitem[\protect\citeauthoryear{{Dong} et~al.,}{{Dong} et~al.}{2021}]{Dong+21}
{Dong} D.~Z.,  et~al., 2021, \mn@doi [Science] {10.1126/science.abg6037}, \href
  {https://ui.adsabs.harvard.edu/abs/2021Sci...373.1125D} {373, 1125}

\bibitem[\protect\citeauthoryear{{Eggleton}}{{Eggleton}}{1983}]{Eggleton1983}
{Eggleton} P.~P.,  1983, \mn@doi [\apj] {10.1086/160960}, \href
  {https://ui.adsabs.harvard.edu/abs/1983ApJ...268..368E} {268, 368}

\bibitem[\protect\citeauthoryear{{Eldridge}, {Izzard}  \& {Tout}}{{Eldridge}
  et~al.}{2008}]{Eldridge+08}
{Eldridge} J.~J.,  {Izzard} R.~G.,   {Tout} C.~A.,  2008, \mn@doi [\mnras]
  {10.1111/j.1365-2966.2007.12738.x}, \href
  {https://ui.adsabs.harvard.edu/abs/2008MNRAS.384.1109E} {384, 1109}

\bibitem[\protect\citeauthoryear{{Eldridge}, {Fraser}, {Maund}  \&
  {Smartt}}{{Eldridge} et~al.}{2015}]{Eldridge+15}
{Eldridge} J.~J.,  {Fraser} M.,  {Maund} J.~R.,   {Smartt} S.~J.,  2015,
  \mn@doi [\mnras] {10.1093/mnras/stu2197}, \href
  {https://ui.adsabs.harvard.edu/abs/2015MNRAS.446.2689E} {446, 2689}

\bibitem[\protect\citeauthoryear{{Fragos}, {Andrews}, {Ramirez-Ruiz}, {Meynet},
  {Kalogera}, {Taam}  \& {Zezas}}{{Fragos} et~al.}{2019}]{Fragos+2019}
{Fragos} T.,  {Andrews} J.~J.,  {Ramirez-Ruiz} E.,  {Meynet} G.,  {Kalogera}
  V.,  {Taam} R.~E.,   {Zezas} A.,  2019, \mn@doi [\apjl]
  {10.3847/2041-8213/ab40d1}, \href
  {https://ui.adsabs.harvard.edu/abs/2019ApJ...883L..45F} {883, L45}

\bibitem[\protect\citeauthoryear{{Gilkis}, {Soker}  \& {Kashi}}{{Gilkis}
  et~al.}{2019}]{Gilkis+19}
{Gilkis} A.,  {Soker} N.,   {Kashi} A.,  2019, \mn@doi [\mnras]
  {10.1093/mnras/sty3008}, \href
  {https://ui.adsabs.harvard.edu/abs/2019MNRAS.482.4233G} {482, 4233}

\bibitem[\protect\citeauthoryear{{Gingold} \& {Monaghan}}{{Gingold} \&
  {Monaghan}}{1977}]{Gingold+Monaghan77}
{Gingold} R.~A.,  {Monaghan} J.~J.,  1977, \mn@doi [\mnras]
  {10.1093/mnras/181.3.375}, \href
  {https://ui.adsabs.harvard.edu/abs/1977MNRAS.181..375G} {181, 375}

\bibitem[\protect\citeauthoryear{{Glanz} \& {Perets}}{{Glanz} \&
  {Perets}}{2021}]{Glanz+21}
{Glanz} H.,  {Perets} H.~B.,  2021, \mn@doi [\mnras] {10.1093/mnras/stab2291},
  \href {https://ui.adsabs.harvard.edu/abs/2021MNRAS.507.2659G} {507, 2659}

\bibitem[\protect\citeauthoryear{{Grichener} \& {Soker}}{{Grichener} \&
  {Soker}}{2019}]{Grichener&Soker19}
{Grichener} A.,  {Soker} N.,  2019, \mn@doi [\apj] {10.3847/1538-4357/ab1d5d},
  \href {https://ui.adsabs.harvard.edu/abs/2019ApJ...878...24G} {878, 24}

\bibitem[\protect\citeauthoryear{{Grichener}, {Sabach}  \& {Soker}}{{Grichener}
  et~al.}{2018}]{Grichener+18}
{Grichener} A.,  {Sabach} E.,   {Soker} N.,  2018, \mn@doi [\mnras]
  {10.1093/mnras/sty1178}, \href
  {https://ui.adsabs.harvard.edu/abs/2018MNRAS.478.1818G} {478, 1818}

\bibitem[\protect\citeauthoryear{{Herant}}{{Herant}}{1994}]{Herant1994}
{Herant} M.,  1994, \memsai, \href
  {https://ui.adsabs.harvard.edu/abs/1994MmSAI..65.1013H} {65, 1013}

\bibitem[\protect\citeauthoryear{{Hillel}, {Schreier}  \& {Soker}}{{Hillel}
  et~al.}{2021}]{Hillel+21}
{Hillel} S.,  {Schreier} R.,   {Soker} N.,  2021, arXiv e-prints, \href
  {https://ui.adsabs.harvard.edu/abs/2021arXiv211201459H} {p. arXiv:2112.01459}

\bibitem[\protect\citeauthoryear{{Hirai} \& {Yamada}}{{Hirai} \&
  {Yamada}}{2015}]{Hirai&Yamada15}
{Hirai} R.,  {Yamada} S.,  2015, \mn@doi [\apj] {10.1088/0004-637X/805/2/170},
  \href {https://ui.adsabs.harvard.edu/abs/2015ApJ...805..170H} {805, 170}

\bibitem[\protect\citeauthoryear{{Hirai}, {Sawai}  \& {Yamada}}{{Hirai}
  et~al.}{2014}]{Hirai+14}
{Hirai} R.,  {Sawai} H.,   {Yamada} S.,  2014, \mn@doi [\apj]
  {10.1088/0004-637X/792/1/66}, \href
  {https://ui.adsabs.harvard.edu/abs/2014ApJ...792...66H} {792, 66}

\bibitem[\protect\citeauthoryear{{Hirai}, {Podsiadlowski}  \& {Yamada}}{{Hirai}
  et~al.}{2018}]{Hirai+18}
{Hirai} R.,  {Podsiadlowski} P.,   {Yamada} S.,  2018, \mn@doi [\apj]
  {10.3847/1538-4357/aad6a0}, \href
  {https://ui.adsabs.harvard.edu/abs/2018ApJ...864..119H} {864, 119}

\bibitem[\protect\citeauthoryear{{Hirai}, {Sato}, {Podsiadlowski},
  {Vigna-G{\'o}mez}  \& {Mandel}}{{Hirai} et~al.}{2020}]{Hirai+2020}
{Hirai} R.,  {Sato} T.,  {Podsiadlowski} P.,  {Vigna-G{\'o}mez} A.,   {Mandel}
  I.,  2020, \mn@doi [\mnras] {10.1093/mnras/staa2898}, \href
  {https://ui.adsabs.harvard.edu/abs/2020MNRAS.499.1154H} {499, 1154}

\bibitem[\protect\citeauthoryear{{Iaconi}, {Reichardt}, {Staff}, {De Marco},
  {Passy}, {Price}, {Wurster}  \& {Herwig}}{{Iaconi}
  et~al.}{2017}]{Iaconi+2017}
{Iaconi} R.,  {Reichardt} T.,  {Staff} J.,  {De Marco} O.,  {Passy} J.-C.,
  {Price} D.,  {Wurster} J.,   {Herwig} F.,  2017, \mn@doi [\mnras]
  {10.1093/mnras/stw2377}, \href
  {https://ui.adsabs.harvard.edu/abs/2017MNRAS.464.4028I} {464, 4028}

\bibitem[\protect\citeauthoryear{{Ivanova}}{{Ivanova}}{2018}]{Ivanova18}
{Ivanova} N.,  2018, \mn@doi [\apjl] {10.3847/2041-8213/aac101}, \href
  {https://ui.adsabs.harvard.edu/abs/2018ApJ...858L..24I} {858, L24}

\bibitem[\protect\citeauthoryear{{Ivanova} \& {Nandez}}{{Ivanova} \&
  {Nandez}}{2016}]{Ivanova&Nandez16}
{Ivanova} N.,  {Nandez} J.~L.~A.,  2016, \mn@doi [\mnras]
  {10.1093/mnras/stw1676}, \href
  {https://ui.adsabs.harvard.edu/abs/2016MNRAS.462..362I} {462, 362}

\bibitem[\protect\citeauthoryear{{Ivanova} et~al.,}{{Ivanova}
  et~al.}{2013a}]{Ivanova+2013}
{Ivanova} N.,  et~al., 2013a, \mn@doi [\aapr] {10.1007/s00159-013-0059-2},
  \href {https://ui.adsabs.harvard.edu/abs/2013A&ARv..21...59I} {21, 59}

\bibitem[\protect\citeauthoryear{{Ivanova}, {Justham}, {Avendano Nandez}  \&
  {Lombardi}}{{Ivanova} et~al.}{2013b}]{Ivanova+13b}
{Ivanova} N.,  {Justham} S.,  {Avendano Nandez} J.~L.,   {Lombardi} J.~C.,
  2013b, \mn@doi [Science] {10.1126/science.1225540}, \href
  {https://ui.adsabs.harvard.edu/abs/2013Sci...339..433I} {339, 433}

\bibitem[\protect\citeauthoryear{{Ivanova}, {Justham}  \&
  {Podsiadlowski}}{{Ivanova} et~al.}{2015}]{Ivanova+15}
{Ivanova} N.,  {Justham} S.,   {Podsiadlowski} P.,  2015, \mn@doi [\mnras]
  {10.1093/mnras/stu2582}, \href
  {https://ui.adsabs.harvard.edu/abs/2015MNRAS.447.2181I} {447, 2181}

\bibitem[\protect\citeauthoryear{{Joyce}, {Lairmore}, {Price}, {Mohamed}  \&
  {Reichardt}}{{Joyce} et~al.}{2019}]{Joyce+19}
{Joyce} M.,  {Lairmore} L.,  {Price} D.~J.,  {Mohamed} S.,   {Reichardt} T.,
  2019, \mn@doi [\apj] {10.3847/1538-4357/ab3405}, \href
  {https://ui.adsabs.harvard.edu/abs/2019ApJ...882...63J} {882, 63}

\bibitem[\protect\citeauthoryear{{Kalogera} \& {Webbink}}{{Kalogera} \&
  {Webbink}}{1998}]{Kalogera1998}
{Kalogera} V.,  {Webbink} R.~F.,  1998, \mn@doi [\apj] {10.1086/305085}, \href
  {https://ui.adsabs.harvard.edu/abs/1998ApJ...493..351K} {493, 351}

\bibitem[\protect\citeauthoryear{{Kramer}, {Schneider}, {Ohlmann}, {Geier},
  {Schaffenroth}, {Pakmor}  \& {R{\"o}pke}}{{Kramer} et~al.}{2020}]{Kramer+20}
{Kramer} M.,  {Schneider} F.~R.~N.,  {Ohlmann} S.~T.,  {Geier} S.,
  {Schaffenroth} V.,  {Pakmor} R.,   {R{\"o}pke} F.~K.,  2020, \mn@doi [\aap]
  {10.1051/0004-6361/202038702}, \href
  {https://ui.adsabs.harvard.edu/abs/2020A&A...642A..97K} {642, A97}

\bibitem[\protect\citeauthoryear{{Kruckow}, {Tauris}, {Langer}, {Sz{\'e}csi},
  {Marchant}  \& {Podsiadlowski}}{{Kruckow} et~al.}{2016}]{Kruckow+16}
{Kruckow} M.~U.,  {Tauris} T.~M.,  {Langer} N.,  {Sz{\'e}csi} D.,  {Marchant}
  P.,   {Podsiadlowski} P.,  2016, \mn@doi [\aap]
  {10.1051/0004-6361/201629420}, \href
  {https://ui.adsabs.harvard.edu/abs/2016A&A...596A..58K} {596, A58}

\bibitem[\protect\citeauthoryear{{Kruckow}, {Tauris}, {Langer}, {Kramer}  \&
  {Izzard}}{{Kruckow} et~al.}{2018}]{Kruckow+18}
{Kruckow} M.~U.,  {Tauris} T.~M.,  {Langer} N.,  {Kramer} M.,   {Izzard} R.~G.,
   2018, \mn@doi [\mnras] {10.1093/mnras/sty2190}, \href
  {https://ui.adsabs.harvard.edu/abs/2018MNRAS.481.1908K} {481, 1908}

\bibitem[\protect\citeauthoryear{{Law-Smith} et~al.,}{{Law-Smith}
  et~al.}{2020}]{Law-Smith+20}
{Law-Smith} J. A.~P.,  et~al., 2020, arXiv e-prints, \href
  {https://ui.adsabs.harvard.edu/abs/2020arXiv201106630L} {p. arXiv:2011.06630}

\bibitem[\protect\citeauthoryear{{Liu}, {Tauris}, {R{\"o}pke}, {Moriya},
  {Kruckow}, {Stancliffe}  \& {Izzard}}{{Liu} et~al.}{2015}]{Liu+15}
{Liu} Z.-W.,  {Tauris} T.~M.,  {R{\"o}pke} F.~K.,  {Moriya} T.~J.,  {Kruckow}
  M.,  {Stancliffe} R.~J.,   {Izzard} R.~G.,  2015, \mn@doi [\aap]
  {10.1051/0004-6361/201526757}, \href
  {https://ui.adsabs.harvard.edu/abs/2015A&A...584A..11L} {584, A11}

\bibitem[\protect\citeauthoryear{{Livio} \& {Soker}}{{Livio} \&
  {Soker}}{1988}]{Livio&Soker88}
{Livio} M.,  {Soker} N.,  1988, \mn@doi [\apj] {10.1086/166419}, \href
  {https://ui.adsabs.harvard.edu/abs/1988ApJ...329..764L} {329, 764}

\bibitem[\protect\citeauthoryear{{Loveridge}, {van der Sluys}  \&
  {Kalogera}}{{Loveridge} et~al.}{2011}]{Loveridge+11}
{Loveridge} A.~J.,  {van der Sluys} M.~V.,   {Kalogera} V.,  2011, \mn@doi
  [\apj] {10.1088/0004-637X/743/1/49}, \href
  {https://ui.adsabs.harvard.edu/abs/2011ApJ...743...49L} {743, 49}

\bibitem[\protect\citeauthoryear{{MacLeod} \& {Ramirez-Ruiz}}{{MacLeod} \&
  {Ramirez-Ruiz}}{2015}]{MacLeod&Ramirez-Ruiz2015}
{MacLeod} M.,  {Ramirez-Ruiz} E.,  2015, in APS April Meeting Abstracts. p.
  U2.004

\bibitem[\protect\citeauthoryear{{MacLeod}, {Antoni}, {Murguia-Berthier},
  {Macias}  \& {Ramirez-Ruiz}}{{MacLeod} et~al.}{2017}]{MacLeod+2017}
{MacLeod} M.,  {Antoni} A.,  {Murguia-Berthier} A.,  {Macias} P.,
  {Ramirez-Ruiz} E.,  2017, \mn@doi [\apj] {10.3847/1538-4357/aa6117}, \href
  {https://ui.adsabs.harvard.edu/abs/2017ApJ...838...56M} {838, 56}

\bibitem[\protect\citeauthoryear{{Mandel} \& {Broekgaarden}}{{Mandel} \&
  {Broekgaarden}}{2021}]{Mandel+Broekgaarden21}
{Mandel} I.,  {Broekgaarden} F.~S.,  2021, arXiv e-prints, \href
  {https://ui.adsabs.harvard.edu/abs/2021arXiv210714239M} {p. arXiv:2107.14239}

\bibitem[\protect\citeauthoryear{{Meyer} \& {Meyer-Hofmeister}}{{Meyer} \&
  {Meyer-Hofmeister}}{1979}]{Meyer+Meyer-Hofmeister79}
{Meyer} F.,  {Meyer-Hofmeister} E.,  1979, \aap, \href
  {https://ui.adsabs.harvard.edu/abs/1979A&A....78..167M} {78, 167}

\bibitem[\protect\citeauthoryear{{Monaghan} \& {Lattanzio}}{{Monaghan} \&
  {Lattanzio}}{1985}]{MonaghanLattanzio1983}
{Monaghan} J.~J.,  {Lattanzio} J.~C.,  1985, \aap, \href
  {https://ui.adsabs.harvard.edu/abs/1985A&A...149..135M} {149, 135}

\bibitem[\protect\citeauthoryear{{Moreno}, {Schneider}, {Roepke}, {Ohlmann},
  {Pakmor}, {Podsiadlowski}  \& {Sand}}{{Moreno} et~al.}{2021}]{Moreno+21}
{Moreno} M.~M.,  {Schneider} F. R.~N.,  {Roepke} F.~K.,  {Ohlmann} S.~T.,
  {Pakmor} R.,  {Podsiadlowski} P.,   {Sand} C.,  2021, arXiv e-prints, \href
  {https://ui.adsabs.harvard.edu/abs/2021arXiv211112112M} {p. arXiv:2111.12112}

\bibitem[\protect\citeauthoryear{{Nandez}, {Ivanova}  \& {Lombardi}}{{Nandez}
  et~al.}{2014}]{Nandez+14}
{Nandez} J.~L.~A.,  {Ivanova} N.,   {Lombardi} J.~C. J.,  2014, \mn@doi [\apj]
  {10.1088/0004-637X/786/1/39}, \href
  {https://ui.adsabs.harvard.edu/abs/2014ApJ...786...39N} {786, 39}

\bibitem[\protect\citeauthoryear{{Ogata}, {Hirai}  \& {Hijikawa}}{{Ogata}
  et~al.}{2021}]{Ogata+21}
{Ogata} M.,  {Hirai} R.,   {Hijikawa} K.,  2021, \mn@doi [\mnras]
  {10.1093/mnras/stab1439}, \href
  {https://ui.adsabs.harvard.edu/abs/2021MNRAS.505.2485O} {505, 2485}

\bibitem[\protect\citeauthoryear{{Ohlmann}, {R{\"o}pke}, {Pakmor}  \&
  {Springel}}{{Ohlmann} et~al.}{2016}]{Ohlmann+16}
{Ohlmann} S.~T.,  {R{\"o}pke} F.~K.,  {Pakmor} R.,   {Springel} V.,  2016,
  \mn@doi [\apjl] {10.3847/2041-8205/816/1/L9}, \href
  {https://ui.adsabs.harvard.edu/abs/2016ApJ...816L...9O} {816, L9}

\bibitem[\protect\citeauthoryear{{Ohlmann}, {R{\"o}pke}, {Pakmor}  \&
  {Springel}}{{Ohlmann} et~al.}{2017}]{Ohlmann+2017}
{Ohlmann} S.~T.,  {R{\"o}pke} F.~K.,  {Pakmor} R.,   {Springel} V.,  2017,
  \mn@doi [\aap] {10.1051/0004-6361/201629692}, \href
  {https://ui.adsabs.harvard.edu/abs/2017A&A...599A...5O} {599, A5}

\bibitem[\protect\citeauthoryear{{Paczynski}}{{Paczynski}}{1976}]{Paczynski1976}
{Paczynski} B.,  1976, in {Eggleton} P.,  {Mitton} S.,   {Whelan} J.,  eds,
  IAU Symposium Vol. 73, Structure and Evolution of Close Binary Systems. p.~75

\bibitem[\protect\citeauthoryear{{Passy} et~al.,}{{Passy}
  et~al.}{2012}]{Passy+2012}
{Passy} J.-C.,  et~al., 2012, \mn@doi [\apj] {10.1088/0004-637X/744/1/52},
  \href {https://ui.adsabs.harvard.edu/abs/2012ApJ...744...52P} {744, 52}

\bibitem[\protect\citeauthoryear{{Paxton}, {Bildsten}, {Dotter}, {Herwig},
  {Lesaffre}  \& {Timmes}}{{Paxton} et~al.}{2011}]{MESA1}
{Paxton} B.,  {Bildsten} L.,  {Dotter} A.,  {Herwig} F.,  {Lesaffre} P.,
  {Timmes} F.,  2011, \mn@doi [\apjs] {10.1088/0067-0049/192/1/3}, \href
  {https://ui.adsabs.harvard.edu/abs/2011ApJS..192....3P} {192, 3}

\bibitem[\protect\citeauthoryear{{Paxton} et~al.,}{{Paxton}
  et~al.}{2013}]{MESA2}
{Paxton} B.,  et~al., 2013, \mn@doi [\apjs] {10.1088/0067-0049/208/1/4}, \href
  {https://ui.adsabs.harvard.edu/abs/2013ApJS..208....4P} {208, 4}

\bibitem[\protect\citeauthoryear{{Paxton} et~al.,}{{Paxton}
  et~al.}{2015}]{MESA3}
{Paxton} B.,  et~al., 2015, \mn@doi [\apjs] {10.1088/0067-0049/220/1/15}, \href
  {https://ui.adsabs.harvard.edu/abs/2015ApJS..220...15P} {220, 15}

\bibitem[\protect\citeauthoryear{{Paxton} et~al.,}{{Paxton}
  et~al.}{2018}]{MESA4}
{Paxton} B.,  et~al., 2018, \mn@doi [\apjs] {10.3847/1538-4365/aaa5a8}, \href
  {https://ui.adsabs.harvard.edu/abs/2018ApJS..234...34P} {234, 34}

\bibitem[\protect\citeauthoryear{{Paxton} et~al.,}{{Paxton}
  et~al.}{2019}]{MESA5}
{Paxton} B.,  et~al., 2019, \mn@doi [\apjs] {10.3847/1538-4365/ab2241}, \href
  {https://ui.adsabs.harvard.edu/abs/2019ApJS..243...10P} {243, 10}

\bibitem[\protect\citeauthoryear{{Podsiadlowski}, {Joss}  \&
  {Hsu}}{{Podsiadlowski} et~al.}{1992}]{Podsiadlowski+92}
{Podsiadlowski} P.,  {Joss} P.~C.,   {Hsu} J.~J.~L.,  1992, \mn@doi [\apj]
  {10.1086/171341}, \href
  {https://ui.adsabs.harvard.edu/abs/1992ApJ...391..246P} {391, 246}

\bibitem[\protect\citeauthoryear{{Podsiadlowski}, {Cannon}  \&
  {Rees}}{{Podsiadlowski} et~al.}{1995}]{Podsiadlowski+95}
{Podsiadlowski} P.,  {Cannon} R.~C.,   {Rees} M.~J.,  1995, \mn@doi [\mnras]
  {10.1093/mnras/274.2.485}, \href
  {https://ui.adsabs.harvard.edu/abs/1995MNRAS.274..485P} {274, 485}

\bibitem[\protect\citeauthoryear{{Price}}{{Price}}{2007}]{Price07}
{Price} D.~J.,  2007, \mn@doi [\pasa] {10.1071/AS07022}, \href
  {https://ui.adsabs.harvard.edu/abs/2007PASA...24..159P} {24, 159}

\bibitem[\protect\citeauthoryear{{Price} \& {Monaghan}}{{Price} \&
  {Monaghan}}{2007}]{Price+Monaghan2007}
{Price} D.~J.,  {Monaghan} J.~J.,  2007, \mn@doi [\mnras]
  {10.1111/j.1365-2966.2006.11241.x}, \href
  {https://ui.adsabs.harvard.edu/abs/2007MNRAS.374.1347P} {374, 1347}

\bibitem[\protect\citeauthoryear{{Price} et~al.,}{{Price}
  et~al.}{2018}]{Price+18}
{Price} D.~J.,  et~al., 2018, \mn@doi [Publications of the Astronomical Society
  of Australia] {10.1017/pasa.2018.25}, \href
  {https://ui.adsabs.harvard.edu/abs/2018PASA...35...31P} {35, e031}

\bibitem[\protect\citeauthoryear{{Prust} \& {Chang}}{{Prust} \&
  {Chang}}{2019}]{Prust&Change19}
{Prust} L.~J.,  {Chang} P.,  2019, \mn@doi [\mnras] {10.1093/mnras/stz1219},
  \href {https://ui.adsabs.harvard.edu/abs/2019MNRAS.486.5809P} {486, 5809}

\bibitem[\protect\citeauthoryear{{Rasio} \& {Livio}}{{Rasio} \&
  {Livio}}{1996}]{Rasio&Livio1996}
{Rasio} F.~A.,  {Livio} M.,  1996, \mn@doi [\apj] {10.1086/177975}, \href
  {https://ui.adsabs.harvard.edu/abs/1996ApJ...471..366R} {471, 366}

\bibitem[\protect\citeauthoryear{{Reichardt}, {De Marco}, {Iaconi}, {Tout}  \&
  {Price}}{{Reichardt} et~al.}{2019}]{Reichardt+2019}
{Reichardt} T.~A.,  {De Marco} O.,  {Iaconi} R.,  {Tout} C.~A.,   {Price}
  D.~J.,  2019, \mn@doi [\mnras] {10.1093/mnras/sty3485}, \href
  {https://ui.adsabs.harvard.edu/abs/2019MNRAS.484..631R} {484, 631}

\bibitem[\protect\citeauthoryear{{Reichardt}, {De Marco}, {Iaconi}, {Chamandy}
  \& {Price}}{{Reichardt} et~al.}{2020}]{Reichardt+20}
{Reichardt} T.~A.,  {De Marco} O.,  {Iaconi} R.,  {Chamandy} L.,   {Price}
  D.~J.,  2020, \mn@doi [\mnras] {10.1093/mnras/staa937}, \href
  {https://ui.adsabs.harvard.edu/abs/2020MNRAS.494.5333R} {494, 5333}

\bibitem[\protect\citeauthoryear{{Ricker} \& {Taam}}{{Ricker} \&
  {Taam}}{2008}]{Ricker&Taam2008}
{Ricker} P.~M.,  {Taam} R.~E.,  2008, \mn@doi [\apjl] {10.1086/526343}, \href
  {https://ui.adsabs.harvard.edu/abs/2008ApJ...672L..41R} {672, L41}

\bibitem[\protect\citeauthoryear{{Ricker} \& {Taam}}{{Ricker} \&
  {Taam}}{2012}]{Ricker&Taam2012}
{Ricker} P.~M.,  {Taam} R.~E.,  2012, \mn@doi [\apj]
  {10.1088/0004-637X/746/1/74}, \href
  {https://ui.adsabs.harvard.edu/abs/2012ApJ...746...74R} {746, 74}

\bibitem[\protect\citeauthoryear{{Ricker}, {Timmes}, {Taam}  \&
  {Webbink}}{{Ricker} et~al.}{2019}]{Ricker+2018}
{Ricker} P.~M.,  {Timmes} F.~X.,  {Taam} R.~E.,   {Webbink} R.~F.,  2019, in
  {Oskinova} L.~M.,  {Bozzo} E.,  {Bulik} T.,   {Gies} D.~R.,  eds,  IAU
  Symposium Vol. 346, IAU Symposium. pp 449--454 (\mn@eprint {arXiv}
  {1811.03656}), \mn@doi{10.1017/S1743921318007433}

\bibitem[\protect\citeauthoryear{{Rimoldi}, {Portegies Zwart}  \&
  {Rossi}}{{Rimoldi} et~al.}{2016}]{Rimoldi+16}
{Rimoldi} A.,  {Portegies Zwart} S.,   {Rossi} E.~M.,  2016, \mn@doi
  [Computational Astrophysics and Cosmology] {10.1186/s40668-016-0015-4}, \href
  {https://ui.adsabs.harvard.edu/abs/2016ComAC...3....2R} {3, 2}

\bibitem[\protect\citeauthoryear{{Rogers} \& {Nayfonov}}{{Rogers} \&
  {Nayfonov}}{2002}]{Rogers+Nayfonov02}
{Rogers} F.~J.,  {Nayfonov} A.,  2002, \mn@doi [\apj] {10.1086/341894}, \href
  {https://ui.adsabs.harvard.edu/abs/2002ApJ...576.1064R} {576, 1064}

\bibitem[\protect\citeauthoryear{{Rogers}, {Swenson}  \& {Iglesias}}{{Rogers}
  et~al.}{1996}]{Rogers+96}
{Rogers} F.~J.,  {Swenson} F.~J.,   {Iglesias} C.~A.,  1996, \mn@doi [\apj]
  {10.1086/176705}, \href
  {https://ui.adsabs.harvard.edu/abs/1996ApJ...456..902R} {456, 902}

\bibitem[\protect\citeauthoryear{{Sabach}, {Hillel}, {Schreier}  \&
  {Soker}}{{Sabach} et~al.}{2017}]{Sabach+17}
{Sabach} E.,  {Hillel} S.,  {Schreier} R.,   {Soker} N.,  2017, \mn@doi
  [\mnras] {10.1093/mnras/stx2272}, \href
  {https://ui.adsabs.harvard.edu/abs/2017MNRAS.472.4361S} {472, 4361}

\bibitem[\protect\citeauthoryear{{Sand}, {Ohlmann}, {Schneider}, {Pakmor}  \&
  {R{\"o}pke}}{{Sand} et~al.}{2020}]{Sand+20}
{Sand} C.,  {Ohlmann} S.~T.,  {Schneider} F. R.~N.,  {Pakmor} R.,   {R{\"o}pke}
  F.~K.,  2020, \mn@doi [\aap] {10.1051/0004-6361/202038992}, \href
  {https://ui.adsabs.harvard.edu/abs/2020A&A...644A..60S} {644, A60}

\bibitem[\protect\citeauthoryear{{Sandquist}, {Taam}, {Chen}, {Bodenheimer}  \&
  {Burkert}}{{Sandquist} et~al.}{1998}]{Sandquist+1998}
{Sandquist} E.~L.,  {Taam} R.~E.,  {Chen} X.,  {Bodenheimer} P.,   {Burkert}
  A.,  1998, \mn@doi [\apj] {10.1086/305778}, \href
  {https://ui.adsabs.harvard.edu/abs/1998ApJ...500..909S} {500, 909}

\bibitem[\protect\citeauthoryear{{Saumon}, {Chabrier}  \& {van Horn}}{{Saumon}
  et~al.}{1995}]{Saumon+95}
{Saumon} D.,  {Chabrier} G.,   {van Horn} H.~M.,  1995, \mn@doi [\apjs]
  {10.1086/192204}, \href
  {https://ui.adsabs.harvard.edu/abs/1995ApJS...99..713S} {99, 713}

\bibitem[\protect\citeauthoryear{{Schr{\o}der}, {MacLeod}, {Loeb},
  {Vigna-G{\'o}mez}  \& {Mandel}}{{Schr{\o}der} et~al.}{2020}]{Schroder+20}
{Schr{\o}der} S.~L.,  {MacLeod} M.,  {Loeb} A.,  {Vigna-G{\'o}mez} A.,
  {Mandel} I.,  2020, \mn@doi [\apj] {10.3847/1538-4357/ab7014}, \href
  {https://ui.adsabs.harvard.edu/abs/2020ApJ...892...13S} {892, 13}

\bibitem[\protect\citeauthoryear{{Soker} \& {Gilkis}}{{Soker} \&
  {Gilkis}}{2018}]{Soker&Gilkis18}
{Soker} N.,  {Gilkis} A.,  2018, \mn@doi [\mnras] {10.1093/mnras/stx3287},
  \href {https://ui.adsabs.harvard.edu/abs/2018MNRAS.475.1198S} {475, 1198}

\bibitem[\protect\citeauthoryear{{Soker}, {Grichener}  \& {Sabach}}{{Soker}
  et~al.}{2018}]{Soker+18}
{Soker} N.,  {Grichener} A.,   {Sabach} E.,  2018, \mn@doi [\apjl]
  {10.3847/2041-8213/aad736}, \href
  {https://ui.adsabs.harvard.edu/abs/2018ApJ...863L..14S} {863, L14}

\bibitem[\protect\citeauthoryear{{Soker}, {Grichener}  \& {Gilkis}}{{Soker}
  et~al.}{2019}]{Soker+19}
{Soker} N.,  {Grichener} A.,   {Gilkis} A.,  2019, \mn@doi [\mnras]
  {10.1093/mnras/stz364}, \href
  {https://ui.adsabs.harvard.edu/abs/2019MNRAS.484.4972S} {484, 4972}

\bibitem[\protect\citeauthoryear{{Sravan}, {Marchant}  \& {Kalogera}}{{Sravan}
  et~al.}{2019}]{Sravan+19}
{Sravan} N.,  {Marchant} P.,   {Kalogera} V.,  2019, \mn@doi [\apj]
  {10.3847/1538-4357/ab4ad7}, \href
  {https://ui.adsabs.harvard.edu/abs/2019ApJ...885..130S} {885, 130}

\bibitem[\protect\citeauthoryear{{Sun}, {Maund}, {Hirai}, {Crowther}  \&
  {Podsiadlowski}}{{Sun} et~al.}{2020}]{Sun+20}
{Sun} N.-C.,  {Maund} J.~R.,  {Hirai} R.,  {Crowther} P.~A.,   {Podsiadlowski}
  P.,  2020, \mn@doi [\mnras] {10.1093/mnras/stz3431}, \href
  {https://ui.adsabs.harvard.edu/abs/2020MNRAS.491.6000S} {491, 6000}

\bibitem[\protect\citeauthoryear{{Taam}, {Bodenheimer}  \& {Ostriker}}{{Taam}
  et~al.}{1978}]{Taam+78}
{Taam} R.~E.,  {Bodenheimer} P.,   {Ostriker} J.~P.,  1978, \mn@doi [\apj]
  {10.1086/156142}, \href
  {https://ui.adsabs.harvard.edu/abs/1978ApJ...222..269T} {222, 269}

\bibitem[\protect\citeauthoryear{{Tauris} et~al.,}{{Tauris}
  et~al.}{2017}]{Tauris+2017}
{Tauris} T.~M.,  et~al., 2017, \mn@doi [\apj] {10.3847/1538-4357/aa7e89}, \href
  {https://ui.adsabs.harvard.edu/abs/2017ApJ...846..170T} {846, 170}

\bibitem[\protect\citeauthoryear{{Terman}, {Taam}  \& {Hernquist}}{{Terman}
  et~al.}{1995}]{Terman+95}
{Terman} J.~L.,  {Taam} R.~E.,   {Hernquist} L.,  1995, \mn@doi [\apj]
  {10.1086/175702}, \href
  {https://ui.adsabs.harvard.edu/abs/1995ApJ...445..367T} {445, 367}

\bibitem[\protect\citeauthoryear{{Thorne} \& {Zytkow}}{{Thorne} \&
  {Zytkow}}{1977}]{Thorne+Zytkow77}
{Thorne} K.~S.,  {Zytkow} A.~N.,  1977, \mn@doi [\apj] {10.1086/155109}, \href
  {https://ui.adsabs.harvard.edu/abs/1977ApJ...212..832T} {212, 832}

\bibitem[\protect\citeauthoryear{{Tutukov} \& {Yungelson}}{{Tutukov} \&
  {Yungelson}}{1993}]{Tutukov&Yungelson93}
{Tutukov} A.~V.,  {Yungelson} L.~R.,  1993, \azh, \href
  {https://ui.adsabs.harvard.edu/abs/1993AZh....70..812T} {70, 812}

\bibitem[\protect\citeauthoryear{{Vigna-G{\'o}mez} et~al.,}{{Vigna-G{\'o}mez}
  et~al.}{2018}]{Vigna-Gomez+2018}
{Vigna-G{\'o}mez} A.,  et~al., 2018, \mn@doi [\mnras] {10.1093/mnras/sty2463},
  \href {http://adsabs.harvard.edu/abs/2018MNRAS.481.4009V} {481, 4009}

\bibitem[\protect\citeauthoryear{{Vigna-G{\'o}mez} et~al.,}{{Vigna-G{\'o}mez}
  et~al.}{2020}]{Vigna-Gomez+20}
{Vigna-G{\'o}mez} A.,  et~al., 2020, \mn@doi [\pasa] {10.1017/pasa.2020.31},
  \href {https://ui.adsabs.harvard.edu/abs/2020PASA...37...38V} {37, e038}

\bibitem[\protect\citeauthoryear{{Vigna-G{\'o}mez}, {Wassink}, {Klencki},
  {Istrate}, {Nelemans}  \& {Mandel}}{{Vigna-G{\'o}mez}
  et~al.}{2021}]{Vigna-Gomez+21}
{Vigna-G{\'o}mez} A.,  {Wassink} M.,  {Klencki} J.,  {Istrate} A.,  {Nelemans}
  G.,   {Mandel} I.,  2021, arXiv e-prints, \href
  {https://ui.adsabs.harvard.edu/abs/2021arXiv210714526V} {p. arXiv:2107.14526}

\bibitem[\protect\citeauthoryear{{Webbink}}{{Webbink}}{1984}]{Webbink1984}
{Webbink} R.~F.,  1984, \mn@doi [\apj] {10.1086/161701}, \href
  {https://ui.adsabs.harvard.edu/abs/1984ApJ...277..355W} {277, 355}

\bibitem[\protect\citeauthoryear{{Wilson} \& {Nordhaus}}{{Wilson} \&
  {Nordhaus}}{2019}]{Wilson&Nordhaus19}
{Wilson} E.~C.,  {Nordhaus} J.,  2019, \mn@doi [\mnras] {10.1093/mnras/stz601},
  \href {https://ui.adsabs.harvard.edu/abs/2019MNRAS.485.4492W} {485, 4492}

\bibitem[\protect\citeauthoryear{{Wilson} \& {Nordhaus}}{{Wilson} \&
  {Nordhaus}}{2020}]{Wilson&Nordhaus20}
{Wilson} E.~C.,  {Nordhaus} J.,  2020, \mn@doi [\mnras]
  {10.1093/mnras/staa2088}, \href
  {https://ui.adsabs.harvard.edu/abs/2020MNRAS.497.1895W} {497, 1895}

\bibitem[\protect\citeauthoryear{{Xu} \& {Li}}{{Xu} \& {Li}}{2010}]{Xu&Li2010}
{Xu} X.-J.,  {Li} X.-D.,  2010, \mn@doi [\apj] {10.1088/0004-637X/716/1/114},
  \href {https://ui.adsabs.harvard.edu/abs/2010ApJ...716..114X} {716, 114}

\bibitem[\protect\citeauthoryear{{Yoon}}{{Yoon}}{2015}]{Yoon15}
{Yoon} S.-C.,  2015, \mn@doi [\pasa] {10.1017/pasa.2015.16}, \href
  {https://ui.adsabs.harvard.edu/abs/2015PASA...32...15Y} {32, e015}

\bibitem[\protect\citeauthoryear{{Yoon}, {Woosley}  \& {Langer}}{{Yoon}
  et~al.}{2010}]{Yoon+10}
{Yoon} S.~C.,  {Woosley} S.~E.,   {Langer} N.,  2010, \mn@doi [\apj]
  {10.1088/0004-637X/725/1/940}, \href
  {https://ui.adsabs.harvard.edu/abs/2010ApJ...725..940Y} {725, 940}

\bibitem[\protect\citeauthoryear{{Zapartas} et~al.,}{{Zapartas}
  et~al.}{2019}]{Zapartas+19}
{Zapartas} E.,  et~al., 2019, \mn@doi [\aap] {10.1051/0004-6361/201935854},
  \href {https://ui.adsabs.harvard.edu/abs/2019A&A...631A...5Z} {631, A5}

\bibitem[\protect\citeauthoryear{{de Kool}}{{de Kool}}{1990}]{deKool1990}
{de Kool} M.,  1990, \mn@doi [\apj] {10.1086/168974}, \href
  {https://ui.adsabs.harvard.edu/abs/1990ApJ...358..189D} {358, 189}

\bibitem[\protect\citeauthoryear{{van den Heuvel}}{{van den
  Heuvel}}{2019}]{vandenHeuvel2019}
{van den Heuvel} E. P.~J.,  2019, \mn@doi [IAU Symposium]
  {10.1017/S1743921319001315}, \href
  {https://ui.adsabs.harvard.edu/abs/2019IAUS..346....1V} {346, 1}

\makeatother
\end{thebibliography}

\appendix

\section{Constructing an artificial red supergiant profile} \label{app:shooting}
To obtain the hydrostatic profile of our \ac{RSG} donor, we solve the following three equations:
\begin{align}
	\textrm{Hydrostatic equilibrium: }& \frac{dP}{dm} = - \frac{Gm}{4\pi r^4} - \frac{g_\text{core}(r)}{4\pi r^2},   \label{eq:hydrostaticeq}\\
	\textrm{Mass continuity: }& \frac{dr}{dm} = \frac{1}{4\pi r^2\rho}, \\
	\textrm{Energy equation: }& s(P,\rho) = s_0 = \text{constant}. \label{eq:entropyconstraint}
\end{align}
These equations solve for the core-softened stellar profile, meaning the mass coordinate $m$, which excludes the mass of the point particle core, ranges from zero to $M-m_\text{core}$. In equation (\ref{eq:hydrostaticeq}), $g_\text{core}(r)$ is the gravitational acceleration due to the point particle core, which is weaker than Newtonian gravity (`softened') when within the core's softening radius (see Appendix \ref{app:softening}). Equation (\ref{eq:entropyconstraint}) fixes the entropy profile to a constant value, $s_0$, for the reasons explained in Section \ref{subsec:donor}. We use the following expression for the specific entropy:
\begin{align}
	s(P,\rho) = \frac{k_B}{\mu m_\text{H}} \ln\Bigg[\frac{T(P,\rho)^{3/2}}{\rho}\Bigg] + \frac{4aT(P,\rho)^3}{3\rho}.
	\label{eq:entropy}
\end{align}
The first term is the specific entropy of a classical, monatomic gas up to additive constants, and the second term is the radiation specific entropy. To obtain $T$ as a function of $\rho$ and $P$ in equation (\ref{eq:entropy}), we solve equation (\ref{eq:pres}) numerically.

We solve equations (\ref{eq:hydrostaticeq}) to (\ref{eq:entropyconstraint}) with the following boundary conditions: 
\begin{align}
	r(M-m_\mathrm{core}) &= R, \\
	P(M-m_\mathrm{core}) &= P_\text{surf}, \\
	\frac{d\rho}{d m}\bigg|_{m=0} &= 0, \\
	\frac{dP}{dm}\bigg|_{m=0} &= 0.
\end{align}
We set $R=619$\Rsun and $P_\text{surf} = 300.2$ dyn cm${}^{-2}$, following our \MESA \ac{RSG} model. Setting a low surface pressure prevents the stellar surface from spontaneously expanding after mapping to the simulation domain, which is particularly important for \ac{SPH} as it usually uses a vacuum surface boundary condition. We solve the above equations using a shooting method, varying the central density and $s_0$ to obtain the desired stellar radius, $R$, and surface pressure, $P_\text{surf}$.
\section{The softened gravitational potential of the stellar core and companion} \label{app:softening}

\begin{figure}
	\centering
	\includegraphics[width=\linewidth]{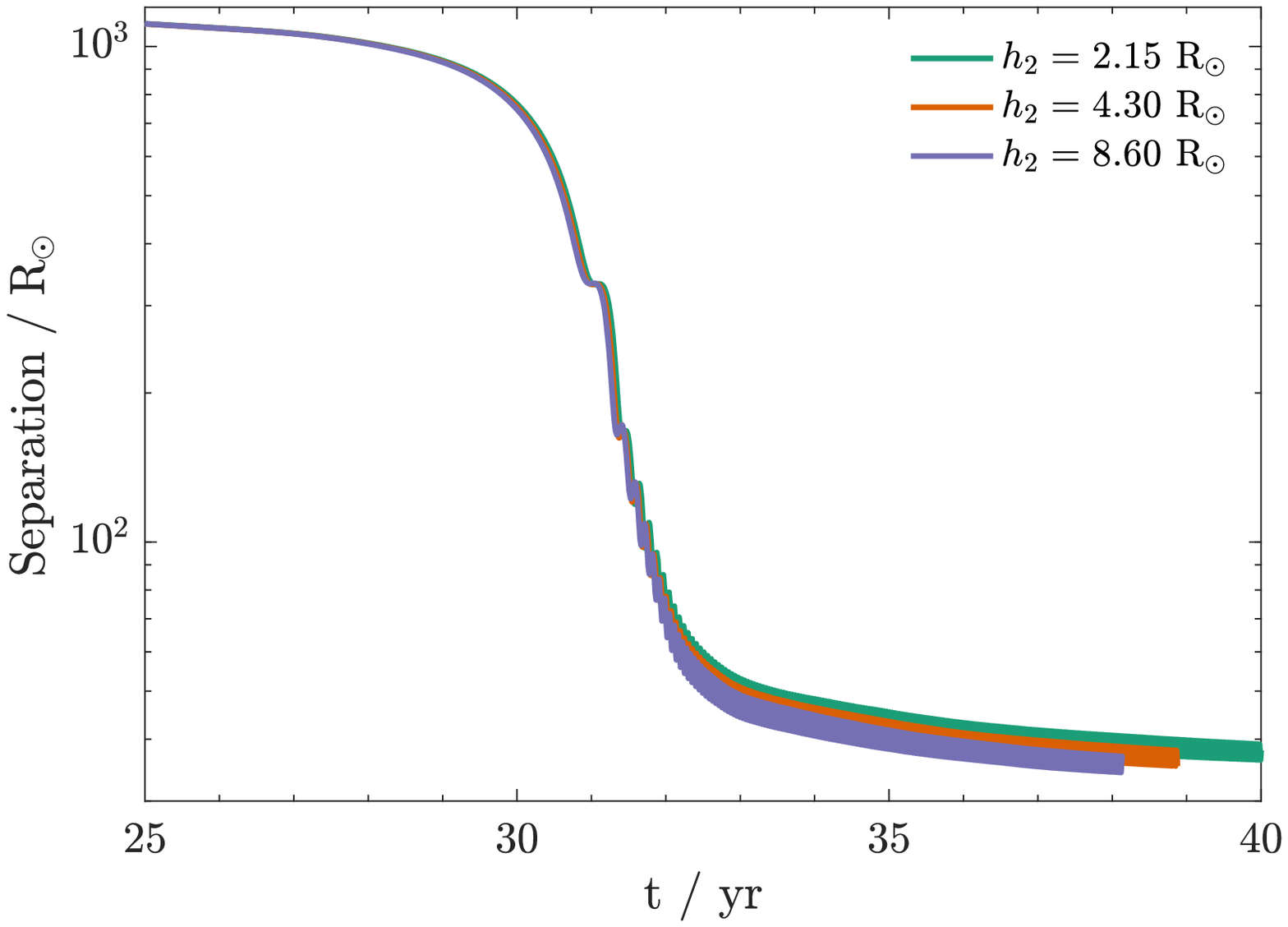}
	\caption{Comparison of the evolution in core-companion separation across simulations with different companion softening lengths, $h_2=2.15\Rsun$ (green), 4.30\Rsun (orange), and 8.60\Rsun (purple).}
	\label{fig:h2_comparison}
\end{figure}

Like most hydrodynamical simulations of \ac{CE} evolution, we do not resolve the dense stellar core of the donor, and instead replace it with a point mass that interacts via a softened gravitational potential. We incorporate this modified gravitational potential into the hydrostatic equilibrium equation when solving for the donor star's structure (equation (\ref{eq:hydrostaticeq})), so that the point mass core is in hydrostatic equilibrium with the gas within its softening radius.

As explained in Section \ref{subsec:donor}, we choose the core mass to be 3.84\Msun, which is the mass coordinate of the base of the convective envelope in the reference \MESA model. The gravitational potential of the point mass is softened with a cubic-spline kernel, whose expression is given in Appendix A of \citet{Price+Monaghan2007}. We choose a softening length of 9.25\Rsun, so that the (unmodified) Newtonian potential is recovered at the base of the convective envelope ($r_\mathrm{core}=18.5\Rsun$, which is twice the softening length).

Like the donor core, the point mass companion also interacts via a cubic-spline-softened potential, but with a softening length of $h_2=2.15$\Rsun. This is consistent with the radius of a 3-4\Msun main-sequence star. The unmodified Newtonian potential is recovered at $r_2=4.30\Rsun$. As mentioned in Section \ref{subsec:sep}, this sets a minimum core-companion separation of $r_\text{core} + r_2 =  22.8\Rsun$, below which the softened region surrounding one stellar core overlaps with the other.

We show that the final separation of a \ac{CE} simulation is not very sensitive to the choice of companion softening length, $h_2$, by comparing the evolution in core-companion separation across simulations with $h_2=2.15$\Rsun (default), 4.30\Rsun, and 8.60\Rsun in Fig. \ref{fig:h2_comparison}. A factor of two increase in $h_2$ causes the semi-major axis during the spiral-in to decrease by $\approx 5$ per cent. This trend might be due to increased feedback from the companion as $h_2$ decreases.
\section{Stellar profile mapping and relaxation}
\label{app:asynchronous}
There are several procedures involved in mapping a 1D stellar profile to a 3D distribution of \ac{SPH} particles: (i) Placing \ac{SPH} particles to achieve the desired initial density profile; (ii) Relaxing the \ac{SPH} particles to ensure the 3D model is in hydrostatic equilibrium; and (iii) Setting a temperature profile consistent with hydrostatic equilibrium with the desired \ac{EoS} in the 3D code.

Determining an initial particle placement amounts to placing particles in a spherical arrangement that matches the desired profile $r(m)$, where $m$ is the radial mass coordinate. In previous versions of \Phantom, this was achieved by placing particles on a uniform Cartesian close-packed lattice cropped to the desired radius, and using the stretch-mapping technique described by \citet{Herant1994}. More sophisticated particle placement schemes have been proposed \citep[e.g.][]{Diehl+15,Joyce+19}. In the previous applications to \ac{CE} simulations \citep{Reichardt+2019,Reichardt+20}, the donor star is subsequently relaxed by evolving it in isolation while including a damping term in the equations of motion \citep[c.f.][]{Gingold+Monaghan77}.

In the present work, we have revised our approach due to three defects with the previous procedure: (i) The stretch-mapped star inherits artefacts associated with the symmetry axes and planes of the lattice that may persist even after relaxing for many dynamical times; (ii) The relaxation procedure is inefficient, as the stellar interior relaxes much more quickly than the surface due to the $r^{3/2}$-scaling of the dynamical time; (iii) The relaxation process works well when a polytropic or barotropic \ac{EoS} is employed. But with more complicated \acp{EoS}, the density profile may evolve away from the desired profile due to evolution in the internal energy.

To remedy these issues, we implement a new {\sc relax-o-matic} procedure in \Phantom to generate initial conditions for our \ac{CE} simulations. The first difference from the previous procedure is the adoption of a purely random (Monte Carlo) initial particle placement matching the desired density profile, since we found that this produces the fastest relaxation to the desired stellar density profile, despite the higher particle noise initially.

The second change is that we implemented our own interpretation of the `Weighted Voronoi Tessellation' technique proposed by \cite{Diehl+15}. These authors essentially proposed evolving the particle distribution using a pairwise force between the particles to find the minimum energy state. We instead compute the usual \ac{SPH} sums to evaluate the acceleration $\bm{a}$ in the equations of motion. But we adopt the key idea from \cite{Diehl+15} to use a particle shifting procedure with no inertia. That is, we shift the position of each particle $i$ at iteration $n$ according to
\begin{equation}
	\bm{x}^{n+1}_i = \bm{x}^n_i + \Delta \bm{x}^n_i,
	\label{eq:shift}
\end{equation}
where 
\begin{equation}
\Delta \bm{x}^n_i \equiv \frac12 \left(C_{\rm cour} h_i/c_{{\rm s}, i}\right)^2 \bm{a}_i.
\end{equation}
The quantity in the parenthesis is the local Courant time step, where $C_{\rm cour}= 0.3$ \citep{MonaghanLattanzio1983}, $h_i$ is the smoothing length, $c_{{\rm s},i}$ is the local sound speed, and $\bm{a}_i$ is the acceleration of the $i$th particle. Importantly, we shift particles \textit{asynchronously}, with independent step sizes determined from their local stability criteria, hence allowing different parts of the star to relax on their local dynamical times.

Finally, we fix the entropy profile during relaxation in order to obtain the desired pressure and density profiles in a non-degenerate way, thus ensuring that the initially noisy particle configuration converges onto the desired equilibrium configuration. Specifically, we evolve the particles assuming an ideal gas \ac{EoS}, but fix the entropy-like quantity, $\tilde{s}(m) := P(m)/\rho(m)^\gamma$, at each iteration of the relaxation procedure, where $\gamma = 5/3$ and $m$ is the mass-ranked coordinate of each particle. We do this by updating the internal energy of each particle to be consistent with $\tilde{s}(m)$ calculated using the original stellar profile. The resulting density profile is shown as the top panel of Fig. \ref{fig:donor_profiles}, which agrees very closely with the intended 1D profile shown as the black dashed line.

At the end of the relaxation procedure, we then use the density and pressure profiles of the relaxed star to calculate the internal energy and temperature profiles using the `true' \ac{EoS} intended for the actual simulation.			
\section{Convergence in initial separation} \label{app:roche-filling}
Due to the large computational expense of our highest resolution simulations with 2M particles, we initiate the binary orbit such that the donor overfills its Roche lobe, with a Roche lobe filling factor of $R/R_L = 1.25$. We demonstrate that this does not affect our conclusions about the role of the assumed \ac{EoS} on the final separation.

\begin{figure}
	\centering
	\includegraphics[width=\linewidth]{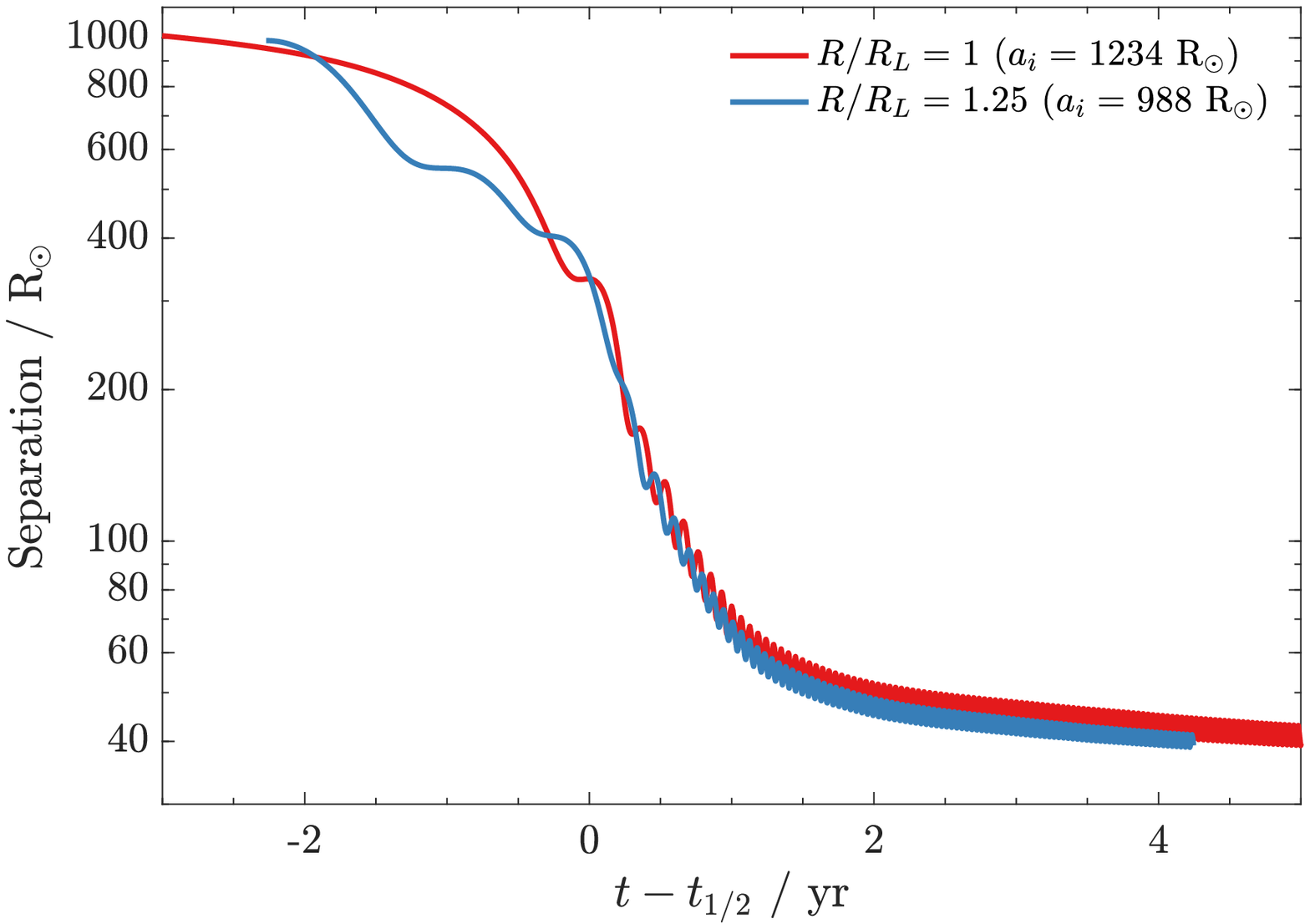}
	\caption{Comparison of the common-envelope inspiral between two simulations that start at different initial separations. The vertical axis shows the core-companion separation, while the horizontal axis marks the time, $t-t_{1/2}$, since the companion has reached half the initial radius of the donor. The red curve corresponds to an initial separation of 1234\Rsun, at which the donor just fills its Roche lobe, while the blue curve corresponds to an initial separation of 988\Rsun, at which the initial radius of the donor is 1.25 times its Roche radius. Both simulations use a 3\Msun companion ($q=0.25$), assume the gas + radiation \ac{EoS}, and are carried out at medium resolution (500k particles).} 
	\label{fig:RL_filling_comparison}
\end{figure}

Figure \ref{fig:RL_filling_comparison} compares the core-companion separations between a simulation initiated at $R/R_L = 1.0$ (red line) and one initiated at $R/R_L = 1.25$ (blue line), both assuming the gas + radiation \ac{EoS} and having medium resolution (500k particles). The companion in the $R/R_L = 1.25$ case reaches half the donor's initial radius in 0.88 years, whereas the $R/R_L = 1.25$ case takes 30 years. Despite this difference, the final separations are very close, with the final separation for the $R/R_L = 1.25$ case being only 6 per cent smaller. This difference between the two simulations may be partly accounted for by the larger energy reservoir in the $R/R_L = 1.0$ case, which has a wider initial separation.


\bsp	
\label{lastpage}
\end{document}